\documentclass[
aps,
prc,
superscriptaddress,
floatfix,
nofootinbib,
twocolumn
]{revtex4-2}

\usepackage{amsmath,amssymb,amsfonts,physics,graphicx,float}
\graphicspath{{pic/}}
\usepackage[setpagesize=false]{hyperref}
\allowdisplaybreaks[4]
\usepackage{subfigure}
\usepackage{ulem}
\usepackage{color}
\definecolor{ar}{rgb}{1.0, 0.01, 0.24}
\definecolor{al}{rgb}{0.82, 0.1, 0.26}
\definecolor{ev}{rgb}{0.56, 0.0, 1.0}

\newcommand{\beq}{\begin{eqnarray}}
\newcommand{\eeq}{\end{eqnarray}}

\begin{document}
\title{ 
Impacts of anomaly on nuclear and neutron star equation of state based on a parity doublet model
}

\author{Bikai Gao}
\email{gaobikai@hken.phys.nagoya-u.ac.jp}
\affiliation{Department of Physics, Nagoya University, Nagoya 464-8602, Japan}

\author{Takuya Minamikawa}
\email{minamikawa@hken.phys.nagoya-u.ac.jp}
\affiliation{Department of Physics, Nagoya University, Nagoya 464-8602, Japan}

\author{Toru Kojo}
\email{torujj@nucl.phys.tohoku.ac.jp}
\affiliation{Department of Physics, Tohoku University, Sendai 980-8578, Japan}
\affiliation{Key Laboratory of Quark and Lepton Physics (MOE) and Institute of Particle Physics,
Central China Normal University, Wuhan 430079, China}

\author{Masayasu Harada}
\email{harada@hken.phys.nagoya-u.ac.jp}
\affiliation{Department of Physics, Nagoya University, Nagoya 464-8602, Japan}
\affiliation{Advanced Science Research Center, Japan Atomic Energy Agency, Tokai 319-1195, Japan}
\affiliation{Kobayashi-Maskawa Institute for the Origin of Particles and the Universe, Nagoya University, Nagoya, 464-8602, Japan}
\date{\today}

\begin{abstract}

We examine the role of the $U(1)_A$ anomaly in a parity doublet model of nucleons which include the chiral variant and invariant masses.
Our model expresses the $U(1)_A$ anomaly by the Kobayashi-Maskawa-'t\,Hooft (KMT) interaction in the mesonic sector.
After examining the roles of the KMT term in vacuum, we discuss its impacts on nuclear equations of state (EOS).
The $U(1)_A$ anomaly increases the masses of the $\eta'$ and $\sigma$ mesons and enhances the chiral symmetry breaking. 
The $U(1)_A$ anomaly enlarges the energy difference between chiral symmetric and symmetry broken vacuum; 
in turn, the chiral restoration at high density adds
a larger energy density (often referred as a bag constant) to EOSs than in the case without the anomaly, leading to softer EOSs.
Including these $U(1)_A$ effects, we update the previously constructed unified equations of state that interpolate the nucleonic EOS at $n_B \le 2n_0$  ($n_{0} = 0.16\, \rm{fm^{-3}}$: nuclear saturation density) and quark EOS at $n_B \ge 5n_0$.
The unified EOS is confronted with the observational constraints on the masses and radii of neutron stars.
The softening of EOSs associated with the $U(1)$ anomaly reduces the overall radii, relaxing the previous constraint on the chiral invariant mass $m_0$.
Including the attractive nonlinear $\rho$-$\omega$ coupling to improved estimates for the slope parameter in the symmetry energy, 
our new estimate is $400\,{\rm MeV} \leq m_0 \leq 700\,{\rm MeV}$, with $m_0$ smaller than our previous estimate  by $\sim 200$ MeV.
\end{abstract}

\maketitle



\section{Introduction}

The chiral $SU(N_f)_L \otimes SU(N_f)_R$ symmetry in quantum chromodynamics (QCD) and its spontaneous symmetry breaking (SSB) 
play the key role in describing the low-energy hadron physics, e.g., the soft pion dynamics and the dynamically generated quark masses \cite{Klevansky:1992qe}.
The chiral condensates, being the order parameters of the chiral SSB, quantify the degree of the chiral SSB, 
and also are useful in characterizing states of matter in QCD at finite temperature and/or density \cite{Baym:2017whm,Hayano:2008vn}.

\begin{figure}[htp]
\centering
\vspace{-1.2cm}
\includegraphics[width=7.5cm]{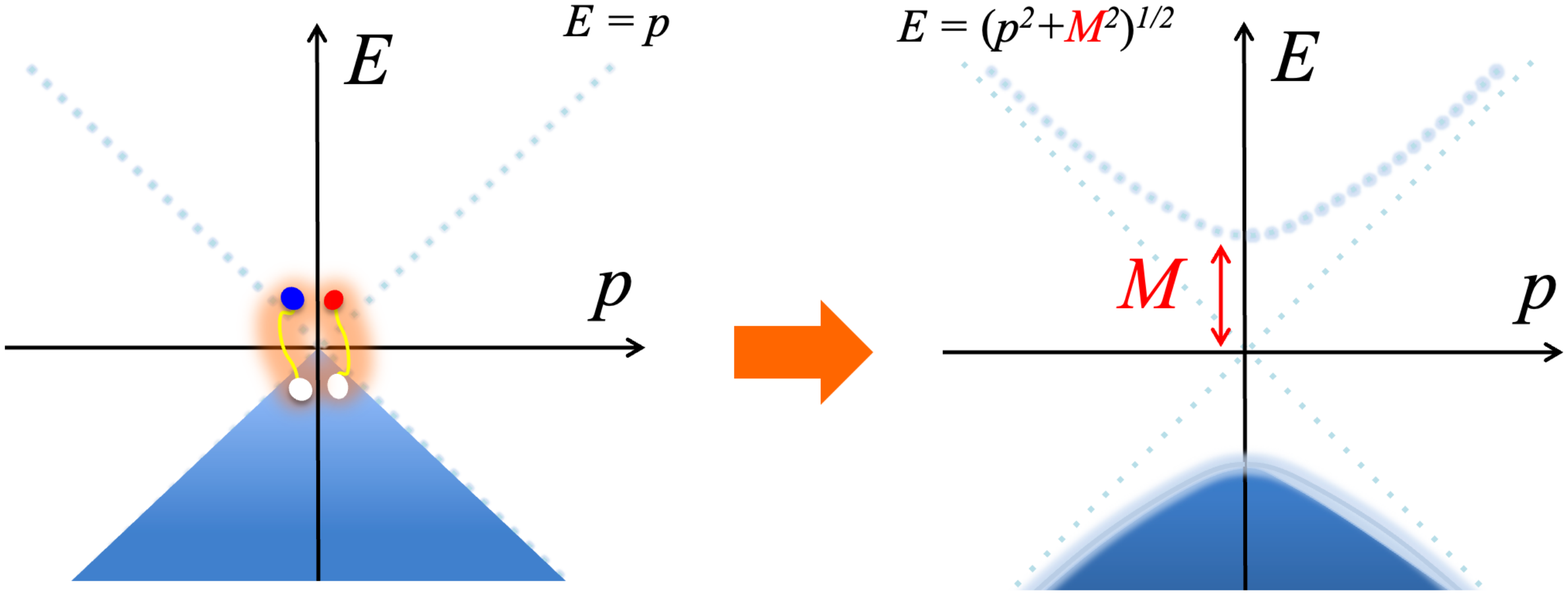}
\vspace{-1.2cm}
\caption{
The Dirac sea in chiral symmetric (left) and symmetry broken (right) phases.
The particle-antiparticle pairs condense to break the chiral symmetry and produce the mass gap $M$.
The mass gap is larger in the presence of the $U(1)_A$ anomaly.
The energy difference in the Dirac sea between the chiral symmetry restored and broken phases defines (a part of) the bag constant. 
}
\label{Dirac_sea}
\end{figure}

In addition to the dynamical SSB, the current quark mass and
the quantum anomaly explicitly break the $U(1)_A$ symmetry and assist the formation of the chiral condensates \cite{PhysRevLett.37.8,THOOFT1986357}.
In this paper we study the impact of the $U(1)_A$ anomaly on the chiral symmetry breaking and examine how it influences nuclear matter equations of state (EOS).
While there are many works on nucleonic EOS emphasizing the importance of in-medium interactions among nucleons, 
in-medium changes of the Dirac sea structure and their impacts on EOS acquire much less attentions.
We argue that the $U(1)_A$ anomaly increases the discrepancy between the chiral symmetry broken and restored phases.
In other words, the anomaly increases the {\it bag constant} associated with the chiral restoration as shown in Fig.\ref{Dirac_sea}. 
In the context of EOS, a larger bag constant adds the energy density but reduces the pressure,
leading to softer EOS.

In the nuclear matter domain, we include the anomaly effects in terms of the Kobayashi-Maskawa-'t\,Hooft (KMT) interactions\cite{Kobayashi:1970ji} 
for a three-flavor mesonic Lagrangian made of scalar and vector mesons.
The KMT interactions relate up-, down-, and strange-quark Dirac sea even before the strangeness appears in a matter.
In fact, the chiral restoration for the up- and down-quark sectors assists the chiral restoration for the strange quark sector, 
possibly changing the masses of hyperons in nuclear matter.
Such structural changes in hyperons are potentially important for matter composition in neutron stars (NSs). 

The baryonic part in this work is treated in a parity doublet model (PDM)\cite{PhysRevD.39.2805,10.1143/PTP.106.873} for nucleons
in which the ordinary nucleon $N(940)$ and its parity partner $N(1535)$ form a doublet. The novel feature of the PDM is that the nucleon masses include not only the conventional chiral variant mass but also invariant mass ($m_0$)
whose existence is supported by the previous lattice QCD simulations\cite{Aarts:2015mma}. 
Accordingly, nucleons in the PDM is less sensitive to the chiral condensate or $\sigma$ fields than in conventional linear $\sigma$ models.
The PDM for vacuum physics has been studied in Refs.\cite{Yamazaki,Minamikawa:2021fln,10.1143/PTP.106.873,PhysRevD.39.2805,Jido:1999hd,Nemoto:1998um,10.1143/PTP.106.873,Nishihara:2015fka,Chen:2009sf,Chen:2010ba,Chen:2011rh}, and its EOS in Refs.\cite{HATSUDA198911,PhysRevC.75.055202,PhysRevC.77.025803,PhysRevC.82.035204,PhysRevD.84.034011,GALLAS201113,PhysRevC.84.045208,PhysRevD.85.054022,PhysRevC.87.015804,PhysRevD.88.105019,PhysRevC.96.025205,PhysRevC.97.045203,PhysRevC.97.065202,refId0,refId01,universe5080180,Motohiro,PhysRevC.100.025205,PhysRevC.103.045205,doi:10.7566/JPSCP.26.024001}.
The parameters of the PDM, coupled to the two-flavor mesonic sector without the $U(1)_A$ anomaly, 
have been tuned to fit the vacuum and the nuclear saturation properties at $n_0$ ($n_0\simeq 0.16\,{\rm fm}^{-3}$: nuclear saturation density).
In this work we retune the parameters including the $U(1)_A$ anomaly.

The key feature of the PDM, in the context of EOS, 
is that a greater $m_0$ leads to weaker $\sigma$ couplings to nucleons,
because a nucleon does not have to acquire its mass entirely from the $\sigma$ fields. 
The couplings to $\omega$ fields are also smaller because, at $n_0$, its repulsive contributions must be balanced with the attractive $\sigma$ contributions.
At densities larger than $n_0$, however, the $\sigma$ fields reduce but the $\omega$ fields increase, 
and these contributions no longer balance;
the repulsive nature of the $\omega$ is directly reflected in the stiffness of EOS.
As a consequence, a larger $m_0$ weakens the $\omega$ fields and softens EOS at supra-nuclear densities.

For applications to NS phenomenology,
 nuclear EOS in the PDM is extrapolated to densities beyond $n_0$.
It has been simply extrapolated  \cite{PhysRevC.100.025205} or combined with a quark model assuming the quark-hadron-crossover \cite{Baym:2017whm,PhysRevC.103.045205,Minamikawa:2021fln,Marczenko:2019trv,Marczenko:2020jma,Masuda:2012kf,Masuda:2012ed}.
In the latter, the PDM EOS is used up to $2n_0$, and interpolate with the quark EOS at $\ge 5n_0$ via polynomial interpolants.
Including the charge neutrality and $\beta$-equilibrium conditions, the unified EOS was confronted with NS constraints
from the existence of two-solar mass ($2M_\odot)$ NSs \cite{NANOGrav:2019jur}
and the gravitational waves from the NS merger event GW170817 \cite{PhysRevLett.119.161101,LIGOScientific:2017ync,LIGOScientific:2018cki}.
Based on the upperbound for the NS radii constraint, 
we previously constrained $m_0$ to rather large values \cite{PhysRevC.103.045205}, $600\,{\rm MeV} \lesssim m_0 \lesssim 900\,{\rm MeV}$.

In this work, we update the constraints by including the $U(1)_A$ anomaly and also include the previously neglected $\rho^2 \omega^2$ terms 
which are usually assumed to be attractive to make EOS softer.
Both effects soften EOS at low densities $\simeq 1$-$2n_0$, leading to smaller NS radii.
As a result, we obtain more relaxed constraints on $m_0$, $400\,{\rm MeV} \lesssim m_0 \lesssim 700\,{\rm MeV}$, reducing the previous range by $\sim 200$-$300$ MeV.
We also add the radius constraint from the PSR J0740+6620 for $2.08\pm 0.07M_\odot$ NS,
$R_{2.08} =12.35 \pm 0.75$ km \cite{Miller:2021qha} 
and $12.39^{+1.30}_{-0.98}$ km \cite{Riley:2021pdl}.

This paper is organized as follows. In 
Sec.\ref{sec:nuclear_matter}, we explain the formulation of our model which based on parity doublet structure. In Sec.\ref{sec:nuclear matter}, we construct EOS in hadronic matter and quark matter separately and the parameters are determined in Sec.\ref{sec:parameters}. Main results of the analysis are shown in Sec.\ref{sec:anomaly} and Sec.\ref{sec:mr}. In Sec.\ref{sec:summary}, we show a summary and discussions.


\section{FORMULATION} \label{sec:nuclear_matter}
In this section, we construct a model of symmetric nuclear matter.


\subsection{Scalar and pseudoscalar mesons} \label{sec:PDM}

We first construct an effective Lagrangian for  scalar and pseudoscalar mesons based on the $SU(3)_{L} \times SU(3)_{R}$ chiral symmetry\cite{Nishihara:2015fka,Chen:2009sf,Chen:2010ba,Chen:2011rh} including the effect of $U(1)_{A}$ anomaly. Quarks transform under $SU(3)_{L} \times SU(3)_{R} \times U(1)_{A}$ symmetry as
\begin{equation}
\begin{aligned}
&q_{L} \rightarrow  e^{-i \theta_{A}} g_{L}q_{L}, \\
&q_{R} \rightarrow  e^{+i \theta_{A}} g_{R}q_{R},
\end{aligned}
\end{equation}
with $g_{L,R} \in SU(3)_{L,R}$ and $\theta_{A}$ being the transformation parameters. 
Accordingly, we assign the $U(1)_{A}$ charge of the left and right handed quarks as $-1$ and $+1$, respectively. 
The chiral representation of the left handed quark is then given by
\begin{equation}
q_{L} :(\mathbf{3}, \mathbf{1})_{-1},
\end{equation}
where these $\mathbf{3}$ and $\mathbf{1}$ in the bracket express the triplet and singlet for $SU(3)_{L}$ symmetry and $SU(3)_{R}$ symmetry, respectively. 
The index indicates the axial charge of the fields. 
On the other hand, the chiral representation of the right-handed quark is given by
\begin{align}
    q_{R}:(\mathbf{1},\mathbf{3})_{+1}.
\end{align}

We introduce a $3\times3$ matrix field $\Phi$ for scalar and pseudoscalar mesons as
\begin{align}
   \Phi_{ij}:(\mathbf{3},\mathbf{\bar{3}})_{-2}.
\end{align}
We adopt the meson part of the Lagrangian as
\begin{align}
    \mathcal{L}_{M}^{\rm{scalar}}=\mathcal{L}_{M}^{\rm{kin}}-V_{M}-V_{\rm{SB}},
\end{align}
where 
\begin{align}
\mathcal{L}_{M}^{\mathrm{kin}}=& \frac{1}{4} \operatorname{tr}\left[\partial_{\mu} \Phi \partial^{\mu} \Phi^{\dagger} \right] ,\\
V_{M}=&-\frac{1}{4} \bar{\mu}^{2} \operatorname{tr}\left[\Phi \Phi^{\dagger}\right]+\frac{1}{8} \lambda_{4} \operatorname{tr}\left[\left(\Phi \Phi^{\dagger}\right)^{2}\right]\nonumber\\
&-\frac{1}{12} \lambda_{6} \operatorname{tr}\left[\left(\Phi \Phi^{\dagger}\right)^{3}\right]+\lambda_{8} \operatorname{tr}\left[\left(\Phi \Phi^{\dagger}\right)^{4}\right]\nonumber\\
&+\lambda_{10} \operatorname{tr}\left[\left(\Phi \Phi^{\dagger}\right)^{5}\right] ,\\
V_{\rm{S B}}=&-\frac{1}{2}c \operatorname{tr}\left[\mathcal{M}^{\dagger} \Phi+\mathcal{M} \Phi^{\dagger}\right],\\
V_{\rm{Anom}}=&-B\left[\operatorname{det}(\Phi)+\operatorname{det}\left(\Phi^{\dagger}\right)\right].
\end{align}
\noindent Here $B$ is the coefficient for the axial anomaly term and $c$ is the coefficient for the explicit chiral symmetry breaking term 
with $\mathcal{M}$ defined as $\mathcal{M}={\rm{diag}}\{m_{u}, m_{d}, m_{s}\}$. 
The above Lagrangian for the meson part is $U(1)_{A}$ invariant except the anomaly term.  
We note that we include only terms with one trace in $V_{M}$, which are expected to be of leading order in $1/N_{c}$ expansion.
Compared with previous model in Ref.\cite{PhysRevC.103.045205}, 
 we not only include the anomaly term but also introduce the $\lambda_{8}$ and $\lambda_{10}$ terms to stablize the potential in the vacuum.\footnote{
If we do not include $\lambda_{8}$ and $\lambda_{10}$ as stabilizers, 
the potential in the vacuum is not bound from below and 
the stationary condition $\partial \mathcal{L}/\partial \sigma_{s}=0$ leads to negative $\sigma_{s}$ values. }

 In this work we use a hadronic model only up to $2n_0$ neglecting hyperons.
Within mean field treatments adopted in this paper, only the diagonal components are kept.
So we reduce $\Phi$ to
\begin{align}
    \Phi=
    \left(
    \begin{matrix}
    M&0\\
    0&\phi_{s}
    \end{matrix}
    \right)_{3\times 3},
\end{align}
where we keep the abstract notation $M$ as a $2\times2$ matrix field to keep track of the $SU(2)_L \times SU(2)_R \times U(1)_A$ structure of our model. 
The meson field under chiral transformation in $ SU(2)$ case is
\begin{align}
	M \rightarrow g_{L}M g_{R}^{\dagger} \,,
\end{align}
where $g_L \in$ $SU(2)_L$ and $g_R \in SU(2)_R$.
Then the reduced Lagrangian is   
\begin{align}
\mathcal{L}_{M}^{\rm{scalar}}=&\frac{1}{4}\left( {\rm tr} \left[ \partial_{\mu}M \partial^\mu M^{\dagger} 
\right]+ \partial_{\mu}\phi_{s}\partial^{\mu}\phi_{s}^{\dagger} \right),\\
V_{M}=&-\frac{1}{4} \bar{\mu}^{2} \left( {\rm tr} \left[MM^{\dagger} \right] + \phi_{s}\phi_{s}^{\dagger}\right)\nonumber\\
&+\frac{1}{8} \lambda_{4} \left( {\rm tr} \left[(MM^{\dagger})^{2}\right]+(\phi_{s}\phi_{s}^{\dagger})^{2}\right)\nonumber\\
&-\frac{1}{12} \lambda_{6} \left({\rm tr} \left[ (MM^{\dagger})^{3} \right]+(\phi_{s}\phi_{s}^{\dagger})^{3}\right)\nonumber\\
&+\lambda_{8} \left( {\rm tr} \left[(MM^{\dagger})^{4}\right]+(\phi_{s}\phi_{s}^{\dagger})^{4}\right)\nonumber\\
&+\lambda_{10}\left({\rm tr} \left[(MM^{\dagger})^{5}\right]+(\phi_{s}\phi_{s}^{\dagger})^{5}\right) ,\\
V_{\rm{S B}}=&-\frac{\, c \,}{2} 
\bigg[
{\rm tr} \left[ \mathcal{M}_{2\times2}
(M+M^{\dagger})\right]+ m_{s} (\phi_{s}+\phi_{s}^{\dagger}) \bigg],\\
V_{\rm{Anom}}=&-B \left[ {\rm{det}}(M)\phi_{s} + {\rm det(}M^{\dagger})\phi_{s}^{\dagger}\right].
\end{align}
where $\mathcal{M}_{2\times2}={\rm diag}\{m_{u},m_{d}\}$.

\subsection{ Nucleon parity doublet and vector mesons   }

While we treat the mesonic sector including three-flavors, we discuss nucleons only up to $2n_0$ where we assume that hyperons do not enter the system.
In the PDM, we assume that nucleons and the chiral partners belong to the representations of $(\mathbf{2},\mathbf{1})_{+1}$ and $ (\mathbf{1}, \mathbf{2})_{-1}$ as 
\begin{align}
    \psi_{1}^{L}: (\mathbf{2},\mathbf{1})_{-1}&,\quad \psi_{1}^{R}: (\mathbf{1},\mathbf{2})_{+1},\\
    \psi_{2}^{L}: (\mathbf{1},\mathbf{2})_{+1}&,\quad \psi_{2}^{R}: (\mathbf{2},\mathbf{1})_{-1},
\end{align}
under $SU(2)_{L}\times SU(2)_{R} \times U(1)_{A} $ symmetry.
In mean field treatments, these fields couple to the two-flavor part in the three-flavor mesonic Lagrangian.
Then the nucleon part  constructed based on the $SU(2)_{R}\times SU(2)_{L} \times U(1)_{A}$  symmetry is given by
\begin{align}
\mathcal{L}_{N}=& \sum_{i=1,2} \bar{\psi}_{i} i \gamma^{\mu} D_{\mu} \psi_{i} \nonumber\\
&-g_{1}\left(\bar{\psi}_{1}^{L}\tau^{2} (M^{\dagger})^{\rm{T}}\tau^{2} \psi_{1}^{R}+\bar{\psi}_{1}^{R} \tau^{2} M^{\rm{T}} \tau^{2} \psi_{1}^{L}\right)\nonumber\\
&-g_{2}\left(\bar{\psi}_{2}^{L} \tau^{2} M^{\rm{T}} \tau^{2} \psi_{2}^{R}+\bar{\psi}_{2}^{R} \tau^{2} (M^{\dagger})^{\rm{T}}\tau^{2} \psi_{2}^{L}\right) \nonumber\\
&-m_{0}\left(\bar{\psi}_{1}^{L} \psi_{2}^{R}-\bar{\psi}_{1}^{R} \psi_{2}^{L}-\bar{\psi}_{2}^{L} \psi_{1}^{R}+\bar{\psi}_{2}^{R} \psi_{1}^{L}\right) ,
\end{align}
where $\tau_i (i=1,2,3) $ are the Pauli matrices.
The couplings $g_{1,2}$ are the Yukawa couplings to the scalar fields for $\psi_{1,2}$ and the origin of the chiral variant masses.
Meanwhile $m_0$ is the chiral invariant mass which originate from the coupling between $\psi_1$ and $\psi_2$.
In the mean field treatment of $\sigma$, the mass spectra are given by
\begin{equation}
m_{\pm}=\sqrt{m_{0}^{2}+\left(\frac{g_{1}+g_{2}}{2}\right)^{2} \sigma^{2}} \mp \frac{g_{1}-g_{2}}{2} \sigma \,,
\end{equation}
where $+$ is for $N(940)$ and $-$ for $N(1535)$ as the mixture of $\psi_1$ and $\psi_2$ fields.
For vanishing $\sigma$, the masses get degenerated, $m_\pm \rightarrow m_0$.

The coupling of vector mesons to nucleons is introduced in the form of the covariant derivatives
%
\begin{equation}
D_{\mu} \psi_{1,2}^{L, R}=\left(\partial_{\mu}-i V_{\mu}\right) \psi_{1,2}^{L, R}.
\end{equation}
with $V_{\mu}$ general external fields including $\omega$ and $\rho$ mesons coupled to baryon number and isospin densities, respectively.

The Lagrangian for vector mesons is based on the hidden local symmetry (HLS) \cite{BANDO1988217,HARADA20031}.
This part is not affected by the $U(1)_A$ anomaly. 
We use the same form as the previous works except addition of the following term
\beq
{\mathcal L}_{\omega \rho} = \lambda_{\omega \rho}\left(g_{\omega} \omega\right)^{2}\left(g_{\rho} \rho\right)^{2} \,,
\eeq
where $\lambda_{\omega \rho}$ is assumed to be positive, meaning the attractive correlation between the $\omega$ and $\rho$ fields.
This term assists the appearance of $\rho$ fields as $\omega$ fields develop.
The $\omega$-$\rho$ correlations play important roles in the symmetry energy, as will be discussed in the following section.

\section{ Nuclear and quark equations of state } 
\label{sec:nuclear matter}

In this section, we construct neutron star matter EOS in both hadronic matter part and quark matter part.

\subsection{Nuclear matter EOS}
\label{sec:PDM matter}

Following Ref.~\cite{PhysRevC.103.045205}, we apply the mean field approximation to the Lagrangian in the last section,
and then calculate the thermodynamic potential in the hadronic matter as
\begin{equation}
\begin{aligned}
\Omega_{\mathrm{PDM}}=& V(\sigma,\sigma_{s})-V\left(\sigma_{0},\sigma_{s0}\right)-\frac{1}{2} m_{\omega}^{2} \omega^{2}-\frac{1}{2} m_{\rho}^{2} \rho^{2} \\
&-\lambda_{\omega \rho}\left(g_{\omega} \omega\right)^{2}\left(g_{\rho} \rho\right)^{2}\\
&-2 \sum_{i=+,-} \sum_{\alpha=p, n} \int^{k_{f}} \frac{\mathrm{d}^3 \mathbf{p}}{(2 \pi)^{3}}\left(\mu_{\alpha}^{*}-E_{\mathrm{p}}^{i}\right).
\end{aligned}
\label{Omega PDM}
\end{equation}
Here $i=+,-$ denote for the parity of nucleons and $E_{\bf p}^{i}=\sqrt{{\bf p}^{2}+m_{i}^{2}}$ is the energy of nucleons with mass $m_{i}$ and momentum ${\bf p}$. 
The crossing term $\omega$-$\rho$ interaction is tuned to adjust the slope parameter, see Sec.\ref{sec:parameters}.
The potential $V(\sigma,\sigma_{s})$ of $\sigma$ and $\sigma_{s}$ mean fields is given by
\begin{align}
 V(\sigma,\sigma_{s})=&-\frac{1}{2} \bar{\mu}^{2}\left(\sigma^{2}+\frac{1}{2} \sigma_{s}^{2}\right)+\frac{1}{4} \lambda_{4}\left(\sigma^{4}+\frac{1}{2} \sigma_{s}^{4}\right) \nonumber\\
 &-\frac{1}{6} \lambda_{6}\left(\sigma^{6}+\frac{1}{2} \sigma_{s}^{6}\right)+\lambda_{8}\left(2 \sigma^{8}+\sigma_{s}^{8}\right) \nonumber \\
&+\lambda_{10}\left(2 \sigma^{10}+\sigma_{s}^{10}\right)-2 B \sigma^{2} \sigma_{s} \nonumber \\
&-\left(2 c m_{u} \sigma+c m_{s} \sigma_{s}\right) \,.
\end{align}

The total thermodynamic potential for the NS is obtained by including the effects of leptons as
\begin{equation}
\Omega_{\mathrm{H}}=\Omega_{\mathrm{PDM}}+\sum_{l=e, \mu} \Omega_{l},
\end{equation}

\noindent where $\Omega_{l}(l=e,\mu)$ are the thermodynamic potentials for leptons,
\begin{equation}
\Omega_{l}=-2 \int^{k_{F}} \frac{d^{3} \mathbf{p}}{(2 \pi)^{3}}\left(\mu_{l}-E_{\mathbf{p}}^{l}\right).
\end{equation}
The mean fields here are determined by following stationary conditions:
\begin{equation}
0=\frac{\partial \Omega_{\mathrm{H}}}{\partial \sigma}, \quad 0=\frac{\partial \Omega_{\mathrm{H}}}{\partial \omega}, \quad 0=\frac{\partial \Omega_{\mathrm{H}}}{\partial \rho}.
\end{equation}
We also need to impose the $\beta$ equilibrium and the charge neutrality conditions,
\begin{align}
\mu_{e}=\mu_{\mu}=-\mu_{Q} ,\\
\frac{\partial \Omega_{\mathrm{H}}}{\partial \mu_{Q}}=n_{p}-n_{l}=0 \,,
\end{align}
 where $\mu_Q$ is the charge chemical potential. 
We then have the pressure in hadronic matter as
\begin{equation}
P_{\mathrm{H}}=-\Omega_{\mathrm{H}}.
\end{equation}

\subsection{ Quark matter EOS   }
\label{NJL matter}

Following Refs.\cite{Baym:2017whm,Baym:2019iky}, we use the NJL quark model to describe the quark matter. 
The model includes three-flavors and $U(1)_A$ anomaly effects through the quark version of the KMT interaction. 
The coupling constants are chosen to be the Hatsuda-Kunihiro parameters 
which successfully reproduce the hadron phenomenology at low energy \cite{Baym:2017whm, Hatsuda:1994pi}: 
$G\Lambda^{2}=1.835, K\Lambda^{5}=9.29$ with $\Lambda=631.4\, \rm{MeV}$, see the definition below.
The couplings $g_{V}$ and $H$ characterize the strength of the vector repulsion and attractive diquark correlations whose range will be examined later 
when we discuss the NS constraints.

We can then write down the thermodynamic potential as
\begin{equation}
\begin{aligned}
\Omega_{\mathrm{CSC}}
=&\, \Omega_{s}-\Omega_{s}\left[\sigma_{f}=\sigma_{f}^{0}, d_{j}=0, \mu_{q}=0\right] \\
&+\Omega_{c}-\Omega_{c}\left[\sigma_{f}=\sigma_{f}^{0}, d_{j}=0\right],
\end{aligned}
\end{equation}
where 
the subscript 0 is attached for the vacuum values, and
\begin{align}
&\Omega_{s}=-2 \sum_{i=1}^{18} \int^{\Lambda} \frac{d^{3} \mathbf{p}}{(2 \pi)^{3}} \frac{\epsilon_{i}}{2} \label{energy eigenvalue},\\
&\Omega_{c}=\sum_{i}\left(2 G \sigma_{i}^{2}+H d_{i}^{2}\right)-4 K \sigma_{u} \sigma_{d} \sigma_{s}-g_{V} n_{q}^{2},
\end{align}
with $\sigma_{f}$ are the chiral condensates, $d_{j}$ are diquark condensates, and $n_{q}$ is the quark density. 
In Eq.(\ref{energy eigenvalue}), $\epsilon_{i}$ are energy eigenvalues obtained from inverse propagator in Nambu-Gorkov bases
\begin{equation}
S^{-1}(k)=\left(\begin{array}{lc}
\gamma_{\mu} k^{\mu}-\hat{M}+\gamma^{0} \hat{\mu} & \gamma_{5} \sum_{i} \Delta_{i} R_{i} \\
-\gamma_{5} \sum_{i} \Delta_{i}^{*} R_{i} & \gamma_{\mu} k^{\mu}-\hat{M}-\gamma^{0} \hat{\mu}
\end{array}\right),
\end{equation}
where
\begin{equation}
\begin{aligned}
M_{i} &=m_{i}-4 G \sigma_{i}+K\left|\epsilon_{i j k}\right| \sigma_{j} \sigma_{k}, \\
\Delta_{i} &=-2 H d_{i} ,\\
\hat{\mu} &=\mu_{q}-2 g_{V} n_{q}+\mu_{3} \lambda_{3}+\mu_{8} \lambda_{8}+\mu_{Q} Q.
\end{aligned}
\end{equation}
$S^{-1}(k)$ is $72\times72$ matrix in terms of the color,
flavor, spin, and Nambu-Gorkov basis, which has 72 eigenvalues. $M_{u,d,s}$ are the constituent masses of $u, d, s$ quarks and $\Delta_{1,2,3}$ are the gap energies. 
The $\mu_{3,8}$ are the color chemical potentials which will be tuned to achieve the color neutrality. 
The total thermodynamic potential including the effect of leptons is 
\begin{equation}
\Omega_{\mathrm{Q}}=\Omega_{\mathrm{CSC}}+\sum_{l=e, \mu} \Omega_{l}.
\end{equation}
The mean fields are determined from the gap equations,
\begin{equation}
0=\frac{\partial \Omega_{\mathrm{Q}}}{\partial \sigma_{i}}=\frac{\partial \Omega_{\mathrm{Q}}}{\partial d_{i}},
\end{equation}
From the conditions for electromagnetic charge neutrality and color charge neutrality, we have
\begin{equation}
n_{j}=-\frac{\partial \Omega_{\mathrm{Q}}}{\partial \mu_{j}}=0,
\end{equation}
where $j = 3,8, Q$. 
The baryon number density $n_{B}$ is determined as
\begin{equation}
n_{q}=-\frac{\partial \Omega_{\mathrm{Q}}}{\partial \mu_{q}},
\end{equation}
where $\mu_{q}$ is $1/3$ of the baryon number chemical potential. After determined all the values, we obtain the pressure as
\begin{equation}
P_{\mathrm{Q}}=-\Omega_{\mathrm{Q}}.
\end{equation}
\begin{table} [tb]
\centering
	\caption{  {\small Physical inputs in vacuum in unit of MeV.  }  }\label{input: mass}
	\begin{tabular}{cccccccc}
		\hline\hline
		~$m_\pi$~&~ $m_K$ ~&~ $f_\pi$ ~&~ $f_{K}$ ~&~ $m_\omega$ ~&~ $m_\rho$ ~&~ $m_+$ ~&~ $m_-$\\
		\hline
		~140 ~&~494~&~ 92.4 ~&~ 109 ~&~ 783 ~&~ 776 ~&~ 939 ~&~ 1535\\
		\hline\hline
	\end{tabular}
	\vspace{0.3cm}
%
\centering
	\caption{  {\small Saturation properties used to determine the model parameters: the saturation density $n_0$, the binding energy $B_0$, the incompressibility $K_0$, symmetry energy $S_0$ and the slope parameter $L_{0}$.}  }
	\begin{tabular}{ccccc}\hline\hline
	~$n_0$ [fm$^{-3}$] ~& $E_{\rm Bind}$ [MeV] ~& $K_0$ [MeV] ~& $S_0$ [MeV] ~& $L_{0}$ [MeV]~\\
	\hline
	0.16 & 16 & 240 & 31 & 57.7\\
	\hline\hline
	\end{tabular}
	\label{saturation}
	\vspace{0.3cm}
%
    \centering
    \caption{  {\small Values of model parameters determined for several choices of $\lambda_{8}'=\lambda_{8}f_{\pi}^{4}$. 
     When $B=600$ MeV, we only find solutions which satisfy the saturation properties in the range: $0\leq\lambda_{8}^{'}\leq7.05$, here we list the boundary values as a typical example. 
    $\lambda_{8}'=0$ is the minimum boundary and $\lambda_{8}'=7.05$ is the maximum boundary. 
    } 
    }
\begin{tabular}{c|c|c|c}
\hline \hline
&~ $m_{0}=800$ [MeV] ~&~ $\lambda_{8}'=0$ ~&~ $\lambda_{8}'=7.05$~  \\
\hline
  &$g_{1}$ & $6.99$  & $6.99$ \\
 &$g_{2}$ & $13.4$ &  $13.4$ \\
 &$\bar{\mu}^{2} / f_{\pi}^{2}$ & $24.83$  & $56.13$ \\
 &$\lambda_{4}$ & $63.52$  & $188.6$  \\
$B=0$ [MeV] ~& $\lambda_{6} f_{\pi}^{2}$& $45.27$   & $227.82$ \\
&$\lambda_{\omega \rho}$ & $0.52$  & $0.79$\\
&$\lambda_{10}f_{\pi}^{6}$ & $0.44$  & $-0.76$ \\
&$g_{\omega N N}$ & $5.12$  & $4.61$ \\
&$g_{\rho N N}$ & $10.25$  & $10.28$ \\
\hline \hline
  &$g_{1}$ & $6.99$  & $6.99$ \\
 &$g_{2}$ & $13.4$ &  $13.4$ \\
 &$\bar{\mu}^{2} / f_{\pi}^{2}$ & $7.22$  & $44.74$ \\
 &$\lambda_{4}$ & $102.8$  & $241.68$  \\
$B=600$ [MeV] ~& $\lambda_{6} f_{\pi}^{2}$& $66.23$   & $256.37$ \\
&$\lambda_{\omega \rho}$ & $0.66$  & $0.98$\\
&$\lambda_{10}f_{\pi}^{6}$ & $0.44$  & $-0.76$ \\
&$g_{\omega N N}$ & $4.17$  & $3.71$ \\
&$g_{\rho N N}$ & $9.31$  & $9.29$ \\
\hline \hline
\end{tabular}
    \label{tab:my_label}
\end{table}


\section{Parameter determination}
\label{sec:parameters}

\begin{figure}[tbp]
\begin{center}
\includegraphics[width=8cm]{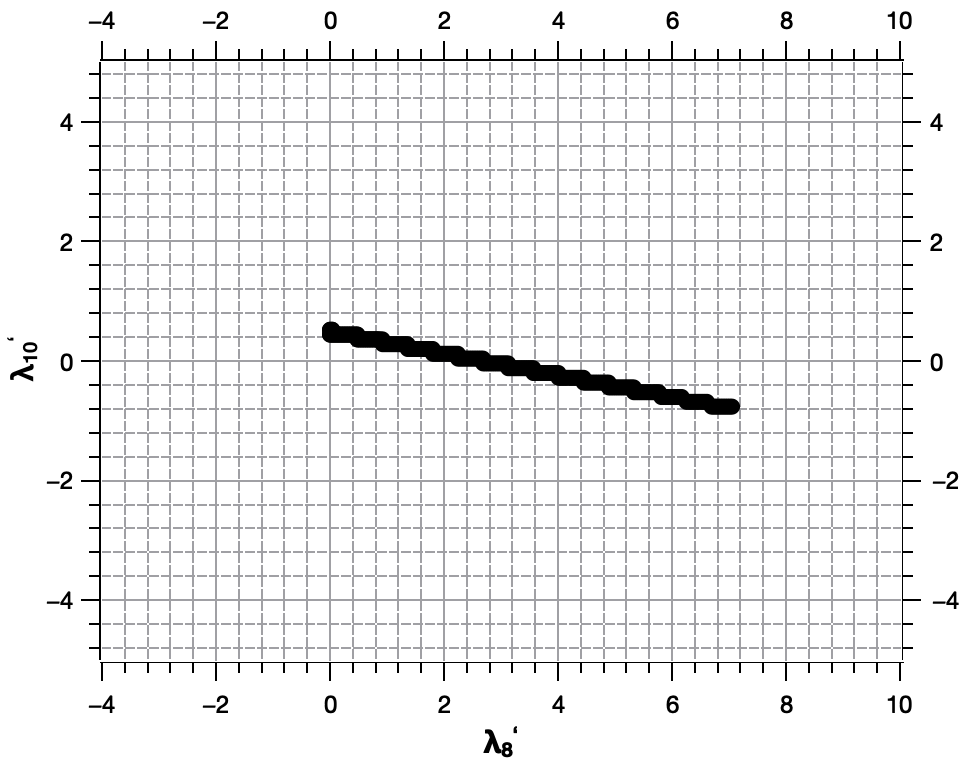}
\caption{Restricted combination of $\lambda_{8}$ and $\lambda_{10}$ after fixing the value of  $\sigma_{s}$ with $m_{0}=800\,\rm{MeV}$. 
$\lambda_{8}^{\prime}=\lambda_{8}f_{\pi}^{4}, \lambda_{10}^{\prime}=\lambda_{10}f_{\pi}^{6}$.
}
\label{lamb8,lamb10}
\end{center}
%
\begin{center}
\includegraphics[width=7cm]{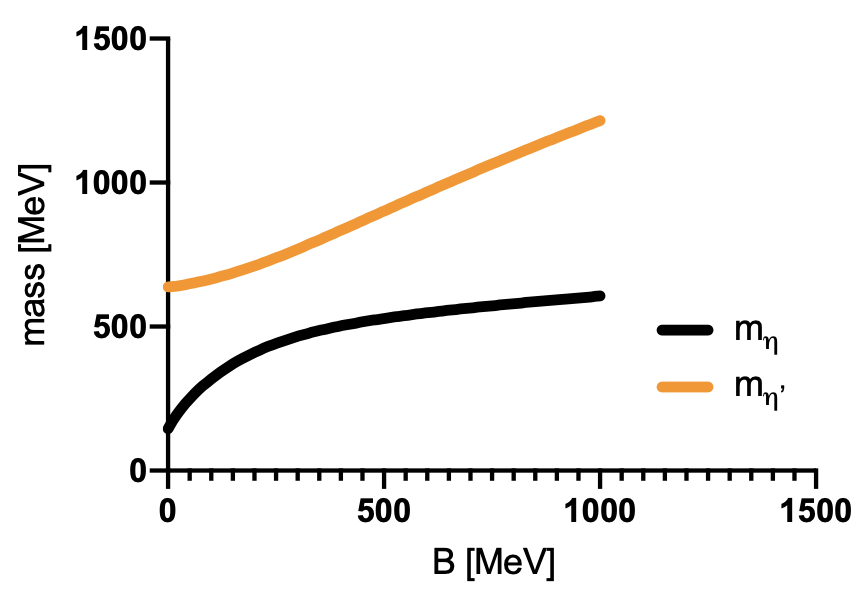}
\caption{ The $B$ dependence of masses of  $\eta$ and $\eta'$ mesons in the vacuum with unit MeV.
}
\label{etap}
\end{center}
\end{figure}

In this section, we determine the parameters in the  PDM
by fitting with the normal nuclear matter properties and the decay constants for different $m_{0}$ (summarized in Table.\ref{input: mass} and Table.\ref{saturation}). 
It is notified that, for $B=\lambda_{8}=\lambda_{10}=0$, the present model is exactly the same as Refs.\cite{Motohiro,PhysRevC.103.045205} 
and the model parameters  can be determined in the same way.  
As in the previous works, we use the vector masses $m_{\rho} =776$ MeV and $m_{\omega} = 783$ MeV.
The parameters $ cm_{u} = cm_d$ and $cm_{s}$ are fixed by the following relations with $f_{\pi}$ and $f_{K}$ given in Table.\ref{input: mass}
\begin{align}
    2cm_{u}=m_{\pi}^{2}f_{\pi}^{2}\,,~~~
    c ( m_{u} + m_{s} ) =m_{K}^{2}f_{K}^{2} \,.
\end{align}
We are left with 11 parameters which will be tuned in the presence of the $U(1)_A$ anomaly.
The mesonic part contains
\beq
  \bar{\mu}^{2},~~ \lambda_{4},~~ \lambda_{6},~~\lambda_{8},~~ \lambda_{10},~~B,~~ \lambda_{\omega \rho} \,,
  \label{eq:mesonic_para}
\eeq
and the nucleonic Lagrangian contains
\beq
m_{0}, ~~g_{1},~~  g_{2},~~g_{\omega NN},~~ g_{\rho NN} \,.
\eeq
In this paper, we treat $m_{0}$ as a given input and then fix the other parameters. 
When we present results for $m_0$ different from the values in this section, 
those results are obtained after retuning the above parameters to achieve the same quality of fitting as in the present section,
unless otherwise stated.

The mesonic part is constrained by the vacuum physics and nuclear saturation properties.
In vacuum, 
the couplings $(g_1, g_2)$, for a given $m_0$, are fixed by demanding $m^{\rm vac}_+ = 939$ MeV and $m^{\rm vac}_- = 1535$ MeV through the relation,
\begin{equation}
m^{\rm vac}_{\pm} =\sqrt{m_{0}^{2}+\left(\frac{g_{1}+g_{2}}{2}\right)^{2} \sigma_0^{2} \,} \mp \frac{\, |g_{1}-g_{2}| \,}{2} \sigma_0 \,.
\end{equation}
where the $\sigma$ fields in vacuum are given by
\begin{align}
    \sigma_{0}=f_{\pi},\quad \sigma_{s0}=f_{K}-\frac{f_{\pi}}{2} \,.
\end{align}
In order to satisfy these relations on $\sigma_0$ and $\sigma_{s0}$,
a proper range of the mesonic parameters in Eq.(\ref{eq:mesonic_para}) must be chosen.

There is still large degeneracy among the mesonic parameters.
We can break the degeneracy by demanding 
the mesonic parameters and ($g_{\omega NN},\, g_{\rho NN},\, \lambda_{\omega \rho}$) to reproduce the saturation properties listed in Table.\ref{saturation}.
Then, we are left with the degeneracy related to the choice of parameters $\lambda_8, \lambda_{10}$, and $B$.
We show the degeneracy related to $\lambda_8$ and $\lambda_{10}$ in Fig.\ref{lamb8,lamb10} by showing the range to reproduce the above-mentioned saturation properties.

Finally, the parameter $B$ is strongly correlated with the $\eta$ and $\eta'$ masses whose experimental values in the vacuum are

\beq
m_{\eta}^{\rm exp} \simeq 547.9\, \rm{MeV}&, \quad m_{\eta^{\prime}}^{\rm exp} \simeq 957.8\, \rm{MeV}.
\eeq

We fix the parameters to reproduce the above-mentioned vacuum and saturation properties for a given $B$. We repeat this procedure while increasing $B$ until the parameters reproduce $\eta$ and $\eta'$ masses correctly. The behaviors of $\eta$ and $\eta'$ masses as functions of $B$ are displayed in Fig. \ref{etap}.
The width attached to the curves reflects the different combinations of $\lambda_{8}$ and $ \lambda_{10}$. 
For $B=600$ MeV, the masses of $\eta^{\prime}$ and $\eta$ are calculated as
\beq
\!\!\!
m^{ {\rm PDM} }_{\eta} = 542 \pm 15\, {\rm MeV},\, 
m^{ {\rm PDM} }_{\eta^{\prime}} = 962 \pm 20 \, {\rm MeV},
\eeq
In this paper, we take $B=600$ MeV as the physical value.

\section{Effect of anomaly in meson sector for hadronic matter}

\begin{figure}[thb]
\centering
\includegraphics[width=7cm]{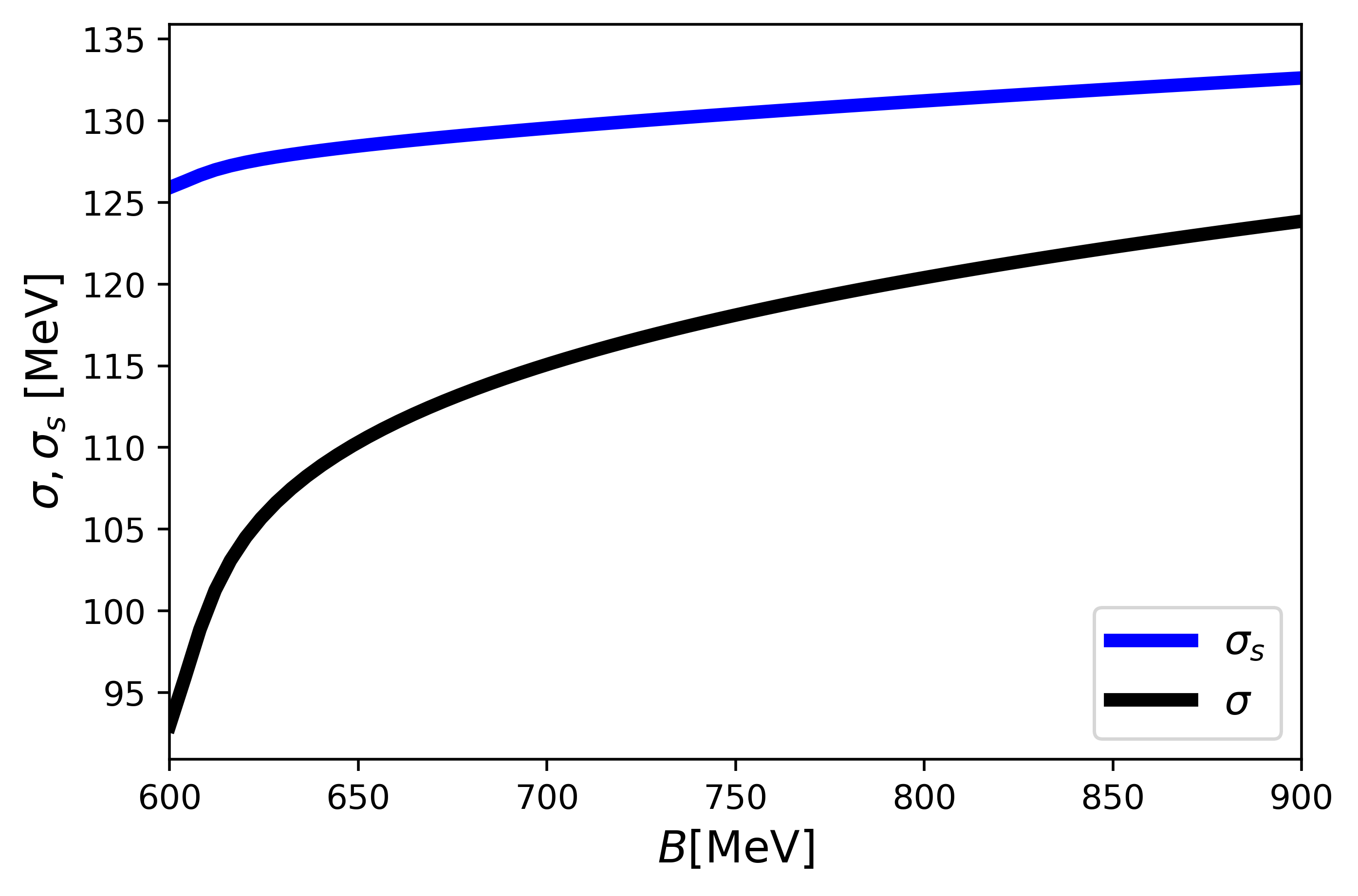}
\caption{
The $B$ dependence of $\sigma$ and $\sigma_{s}$ for $m_{0}=500 \,\rm{MeV}$.  }
\label{vacuum sigma}
\end{figure}

\begin{figure}[thb]
\centering
\includegraphics[width=7cm]{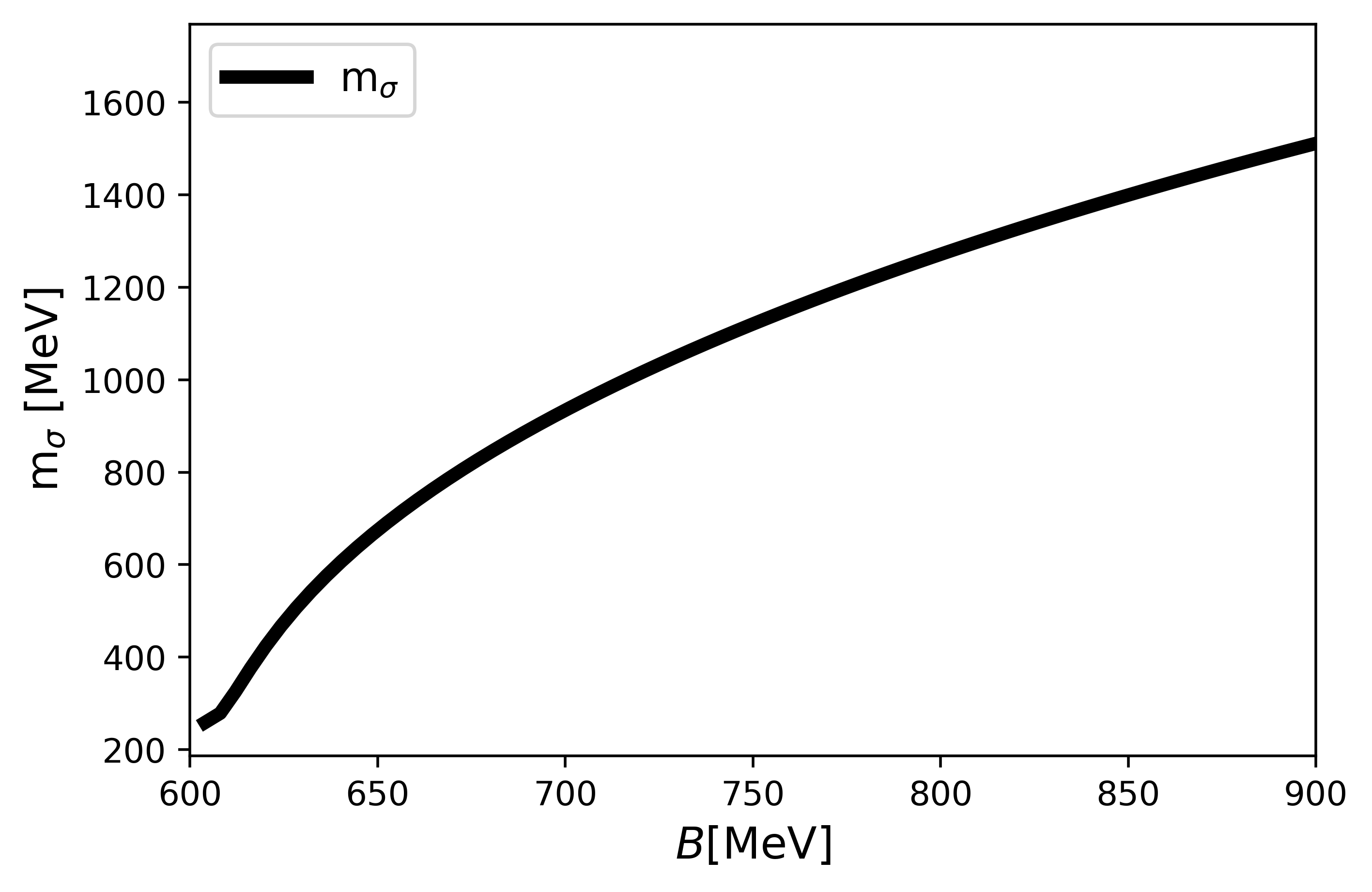}
\caption{
The $B$ dependence of  $m_{\sigma}$ for $m_{0}=500 \,\rm{MeV}$.  }
\label{vacuum sigma_mass}
\end{figure}

To study the effect of the anomaly, we perform linear analysis with respect to the variation of $B$;
we weakly vary the value of $B$ around our physical choice $B=600$ MeV, leaving the other parameters unchanged. 
(Within this linear analysis, the results other than $B=600$ MeV do not satisfy the saturation properties.)

The vacuum value of $\sigma$ and $\sigma_{s}$ change as shown in Fig.\ref{vacuum sigma}. 
The vacuum values of $\sigma$ and $\sigma_{s}$ increase as $B$ does.
This indicates that the anomaly enhances the chiral symmetry breaking, as advertised in the previous sections.
The energy density in vacuum is reduced more by the stronger chiral symmetry breaking. 
When the chiral symmetry is restored, this energy reduction in vacuum is lost, 
and we have to add more energy density or a {\it bag constant} to the EOS in the chiral restored phase.

Another important effect of the anomaly is the increase of $\sigma$ meson mass, as shown in Fig.\ref{vacuum sigma_mass}.
In the context of nuclear forces, the heavier $\sigma$ meson mass reduces the range of attractive force and weakens the overall strength;
this in turn requires weaker repulsive $\omega$ interactions to balance with the $\sigma$ attraction to satisfy the saturation properties.
The resultant reduced repulsion leads to a softer nuclear EOS at supra-saturation densities where $\omega$ dominates over $\sigma$.
In summary, the $U(1)_A$ anomaly effects softens nuclear EOS at supra-saturation densities.

\begin{figure}[tp]
\centering
\includegraphics[width=7cm]{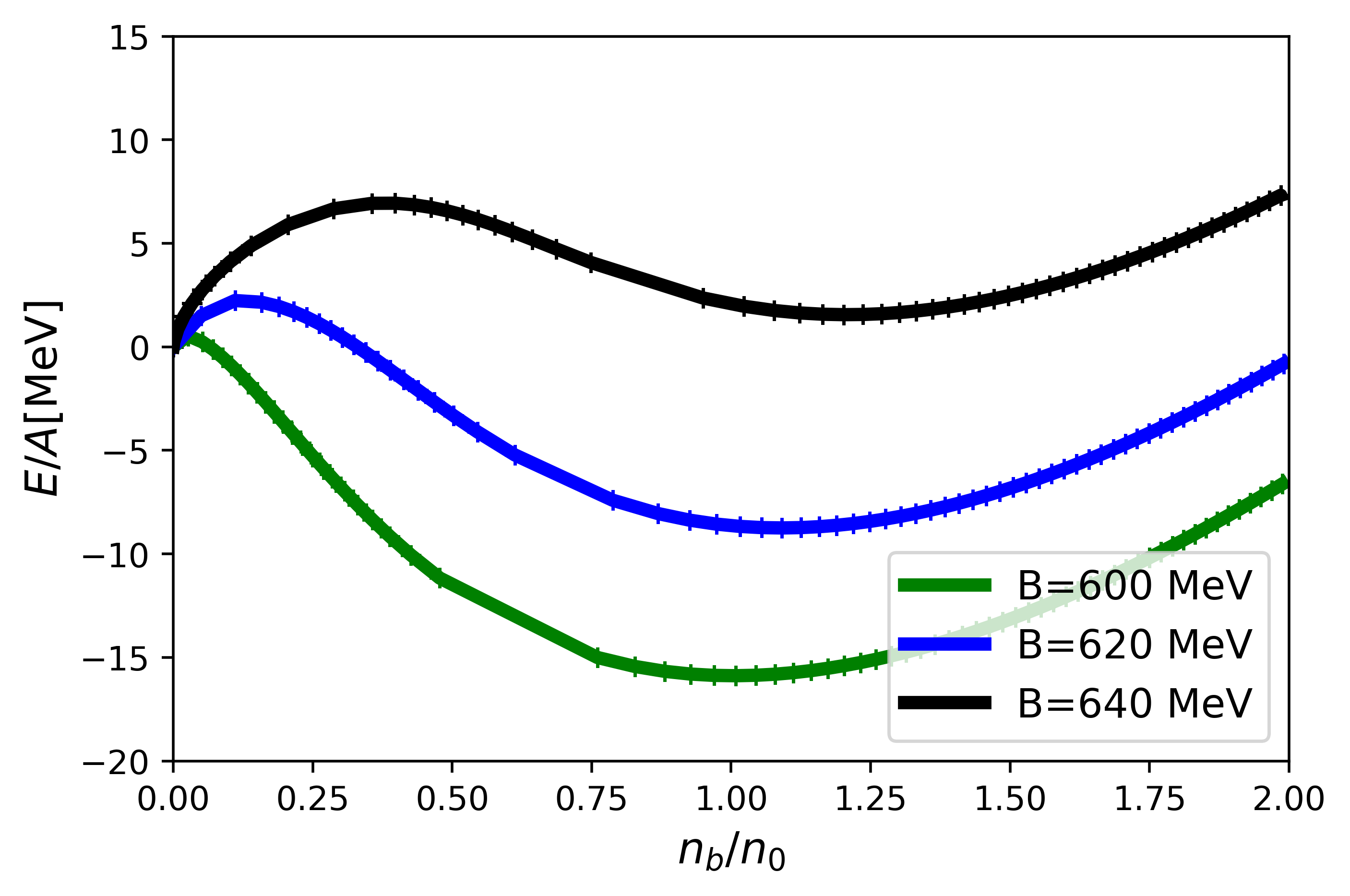}
\caption{Density dependence of the energy for $m_{0}=500\,\rm{MeV}$. 
}
\label{energy-density}
\end{figure}
\begin{figure}[tp]
\centering
\includegraphics[width=7cm]{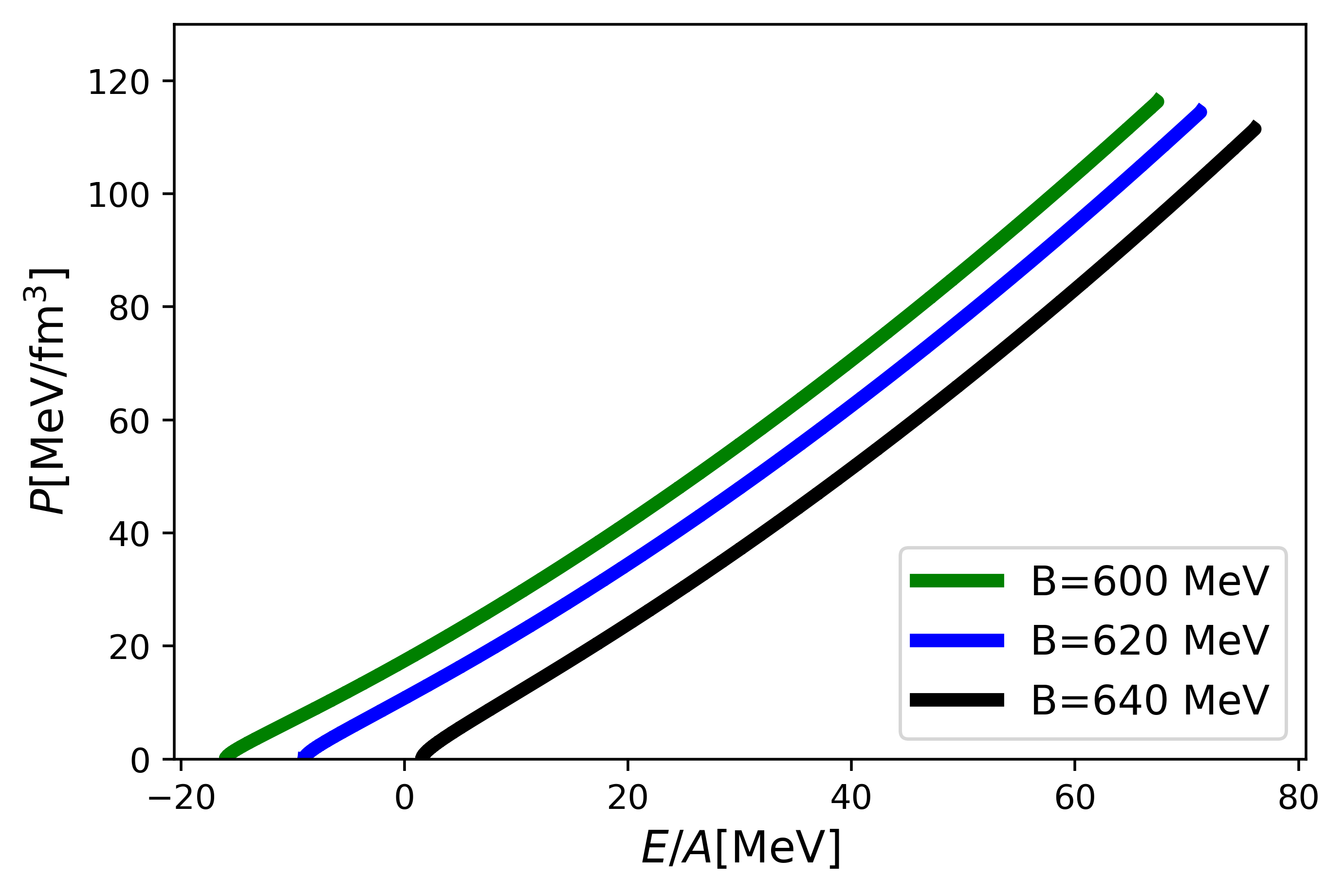}
\caption{The energy dependence of pressure for $m_{0}=500\,\rm{MeV}$. 
}
\label{nb-miuB}
\end{figure}

In Fig.\ref{energy-density}, we show the density dependence of the energy density for $B=580$, $600$ and $620$ MeV with $m_0 = 800$ MeV. 
The energy density overall increases as $B$ does in whole density region, and the saturation points shift to higher densities. 
This can be understood by the competition between the $\sigma$ attraction and $\omega$ repulsion. 
In the present linear analyses, increasing $B$ does not change the vector meson mass but increases the mass of $\sigma$.
As a result, the range of $\sigma$ attraction, $\sim 1/m_\sigma$, decreases as $B$ increases, reducing the attractive contributions to the energy density.
We also show the energy dependence of the pressure in the Fig.\ref{nb-miuB} is obtained through
\begin{align}
    P=\mu_{B}n_{B}- \varepsilon,
\end{align}
which indicates that the effect of anomaly softens the equation of state.


\section{Effect of anomaly in NJL-type model for quark matter}
\label{sec:anomaly}

In the NJL-type model introduced in Sec.~\ref{NJL matter}, the coefficient $K$ represents the strength of anomaly. 
Here we gradually decrease the value of $K$ from $K\Lambda^{5}=9.29$ toward $0$ with fixing other parameters to study the effect of anomaly. For simplicity, we first set  $H=0$ to avoid diquark condensate. 
The chiral condensates in the vacuum have the anomaly dependence as in Fig.\ref{chiral condensate,NJL}.

\begin{figure}[htp]
\includegraphics[width=7cm]{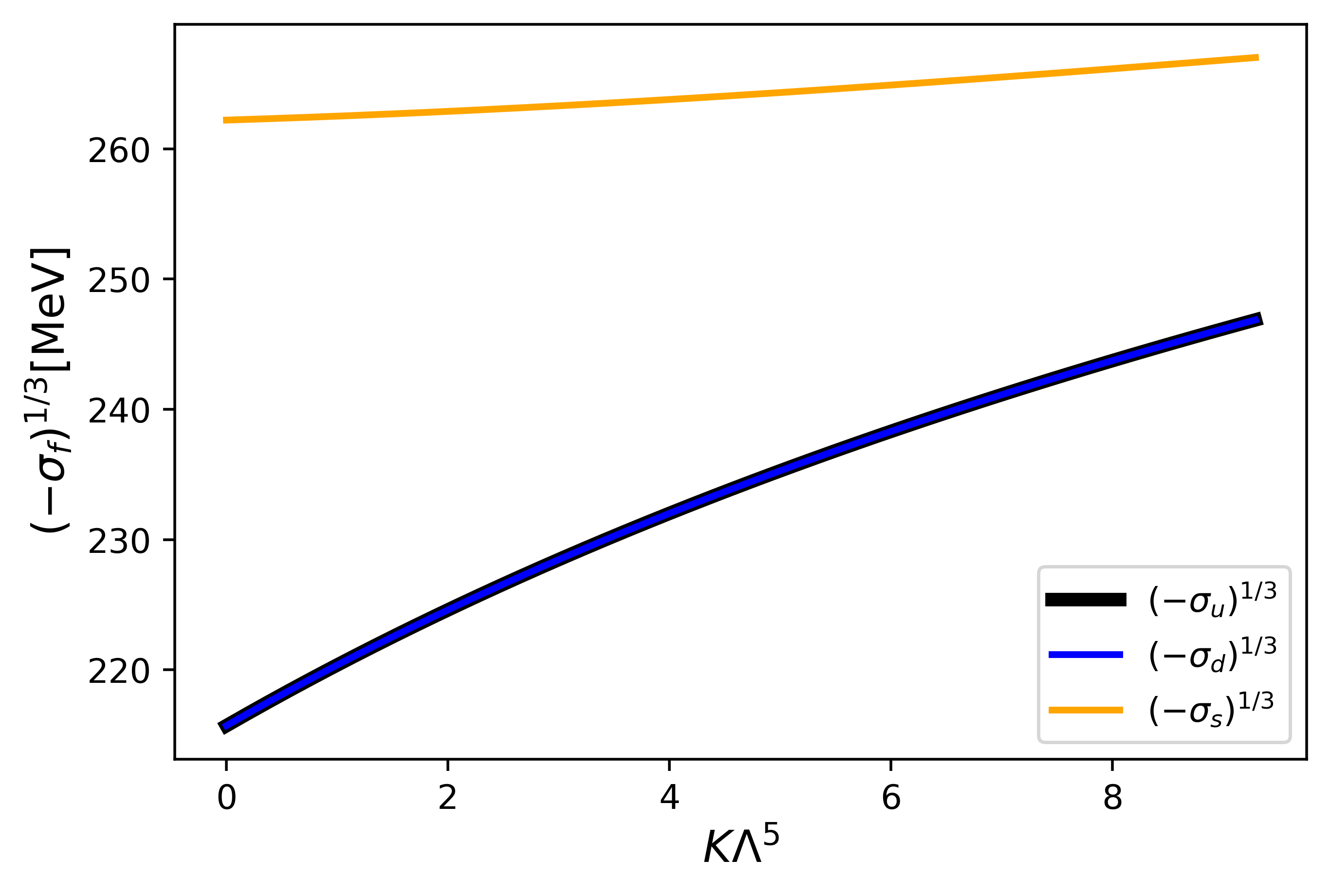}
\caption{
The dependence of chiral condensates on 
 the value of $K$. The horizontal axis shows the value of $K$ normalized as $K \Lambda^5$.
}
\label{chiral condensate,NJL}
\end{figure}

Chiral condensates  in vacuum increase with increasing $K$, which is similar to the PDM. 
This result indicates that the anomaly enhances the chiral symmetry breaking and reduces the ground state energy in vacuum. 

In Fig.\ref{Dirac_sea} we show dispersion relations of quarks in the chiral symmetry broken vacuum (left panel) and in the chiral symmetric vacuum (right panel). 
The approximate chiral symmetry  is spontaneously broken by the chiral condensate, and  quarks of different chiralities are connected with each other. 
Then the condensation opens a gap $M$ in the quark dispersion relation.
As a result, the structure of the Dirac sea is changed to generate  a non-perturbative QCD vacuum. 
The difference in energy density between the chiral symmetric Dirac sea and symmetry-broken Dirac sea defines the bag constant \cite{Kojo:2014rca},
\begin{align}
\varepsilon_{{\rm bag}}=\varepsilon(M_{{\rm eff}}=m_{q})-\varepsilon(M_{{\rm eff}}=M),
\end{align}
where $M_{{\rm eff}}$ is the effective mass of quarks, $m_{q}$ is the bare quark mass and $M$ is the constituent quark mass.

The density dependence of $\varepsilon_{\rm{total}}$ and $\varepsilon_{\rm{bag}}$ are calculated separately as shown in the Fig.\ref{NJL,E-nb,bag} for two cases, 
$K=0$ and $K=9.29/\Lambda^{5}$.
This indicates that
$\varepsilon_{\rm{bag}}>\varepsilon_{\rm{bag}}^{K=0}$ 
at the same density, which implies that the effect of anomaly enhances the bag constant and finally increase the total energy. 

\begin{figure}[tbp]
\centering
\includegraphics[width=7cm]{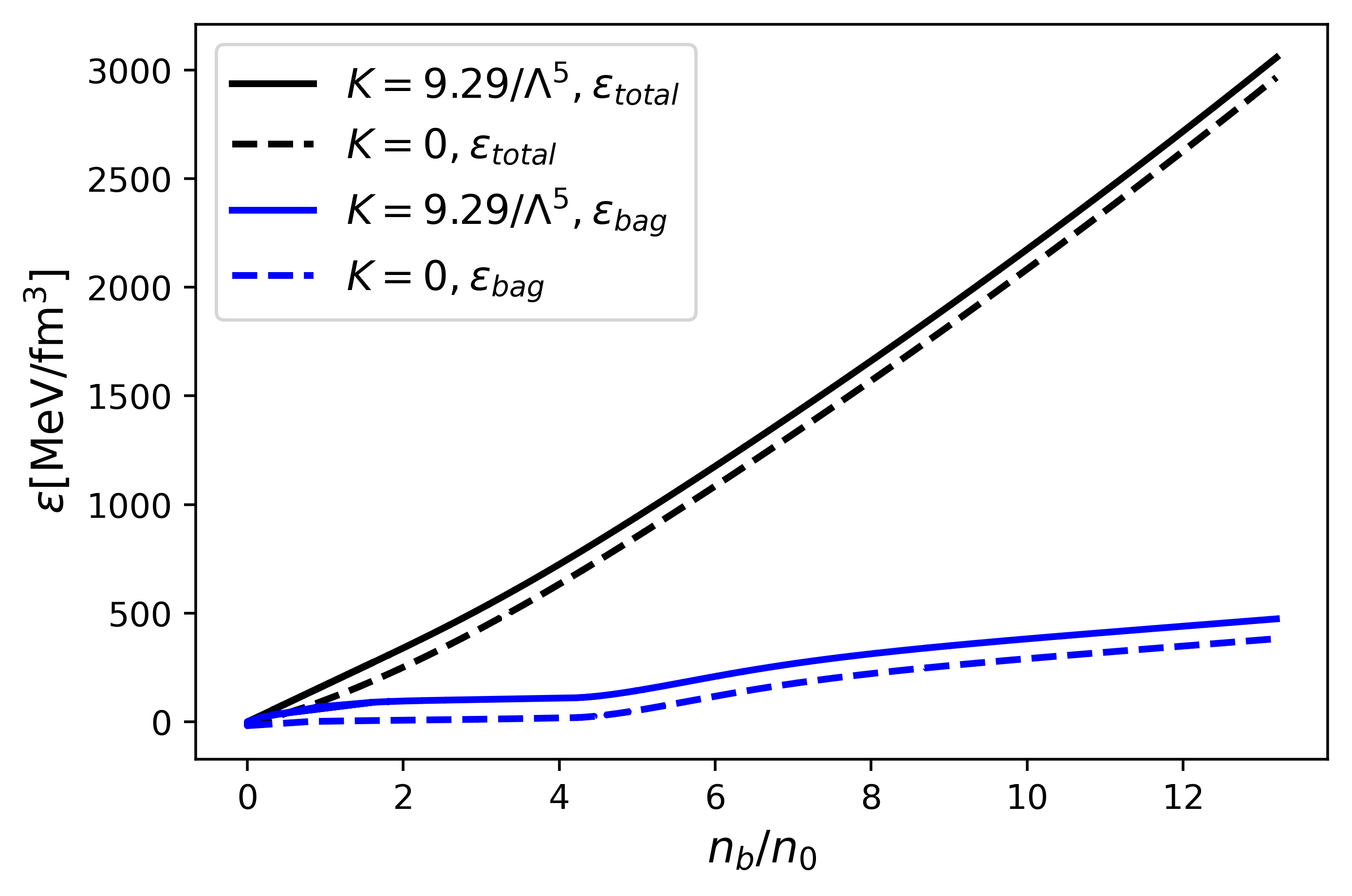}
\caption{The density dependence of $\varepsilon_{\rm{total}}$ and $ \varepsilon_{\rm{bag}}$ with $(H, g_V)/G= (0, 0.1)$. 
}
\label{NJL,E-nb,bag}
\end{figure}

From the analysis of the chiral condensates in the vacuum in Fig.\ref{chiral condensate,NJL}, anomaly effect lowers the ground state energy of the vacuum. 
In Fig.\ref{Dirac_sea}, we show a schematic view of vacuum structure. 
 The released energy after chiral symmetry restoration is larger with anomaly than without it, 
then at the same density $\varepsilon_{\rm{bag}} >\varepsilon_{\rm{bag}}^{K=0}$.

\begin{figure}[tbp]
\centering
\includegraphics[width=7cm]{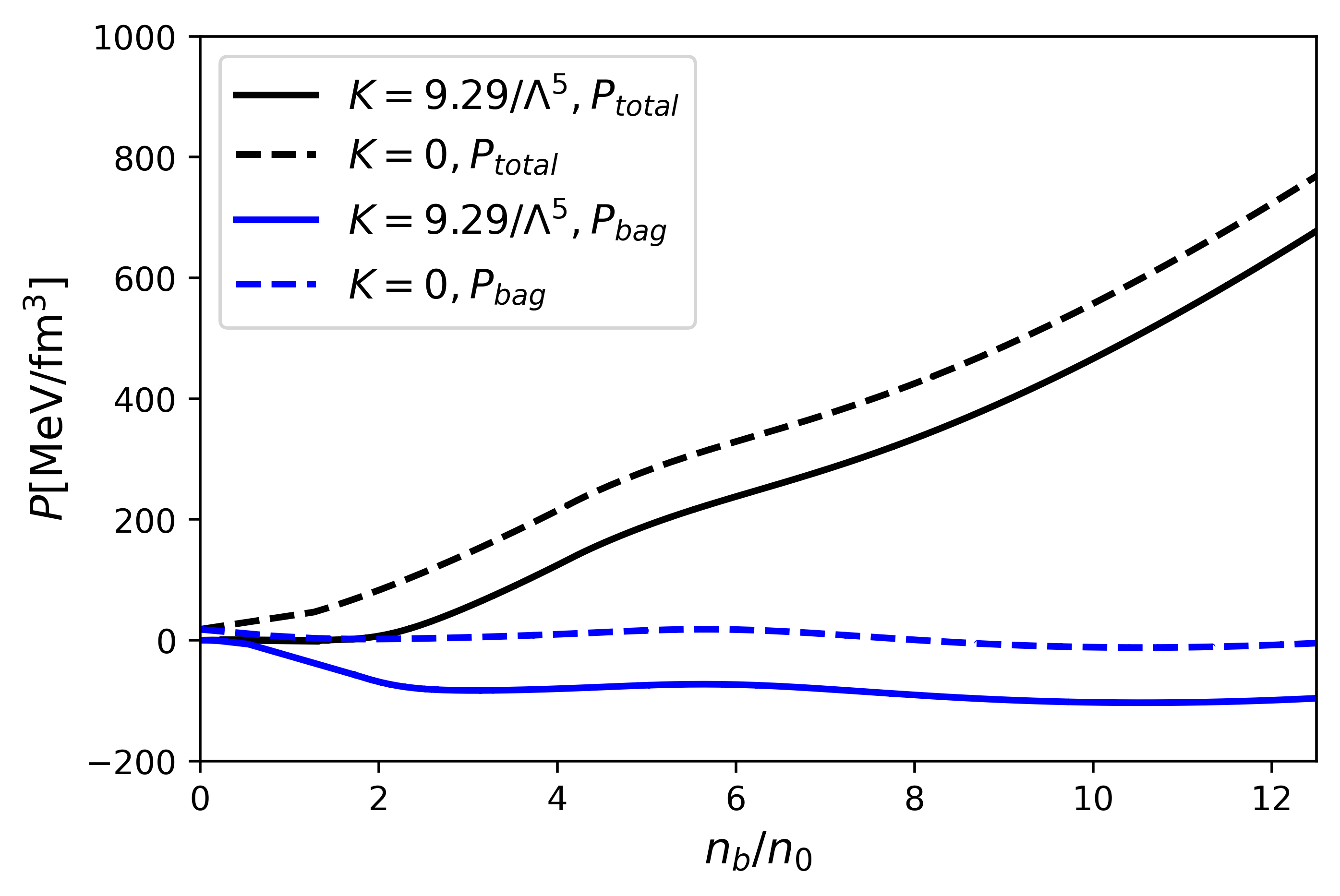}
\caption{Density dependence of $P_{\rm{total}}$ and $ P_{\rm{bag}}$ with $(H, g_V)/G= (0, 0.1)$. 
 }
\label{pressure-density}
\end{figure}

\begin{figure}[htp]
\centering
\includegraphics[width=7cm]{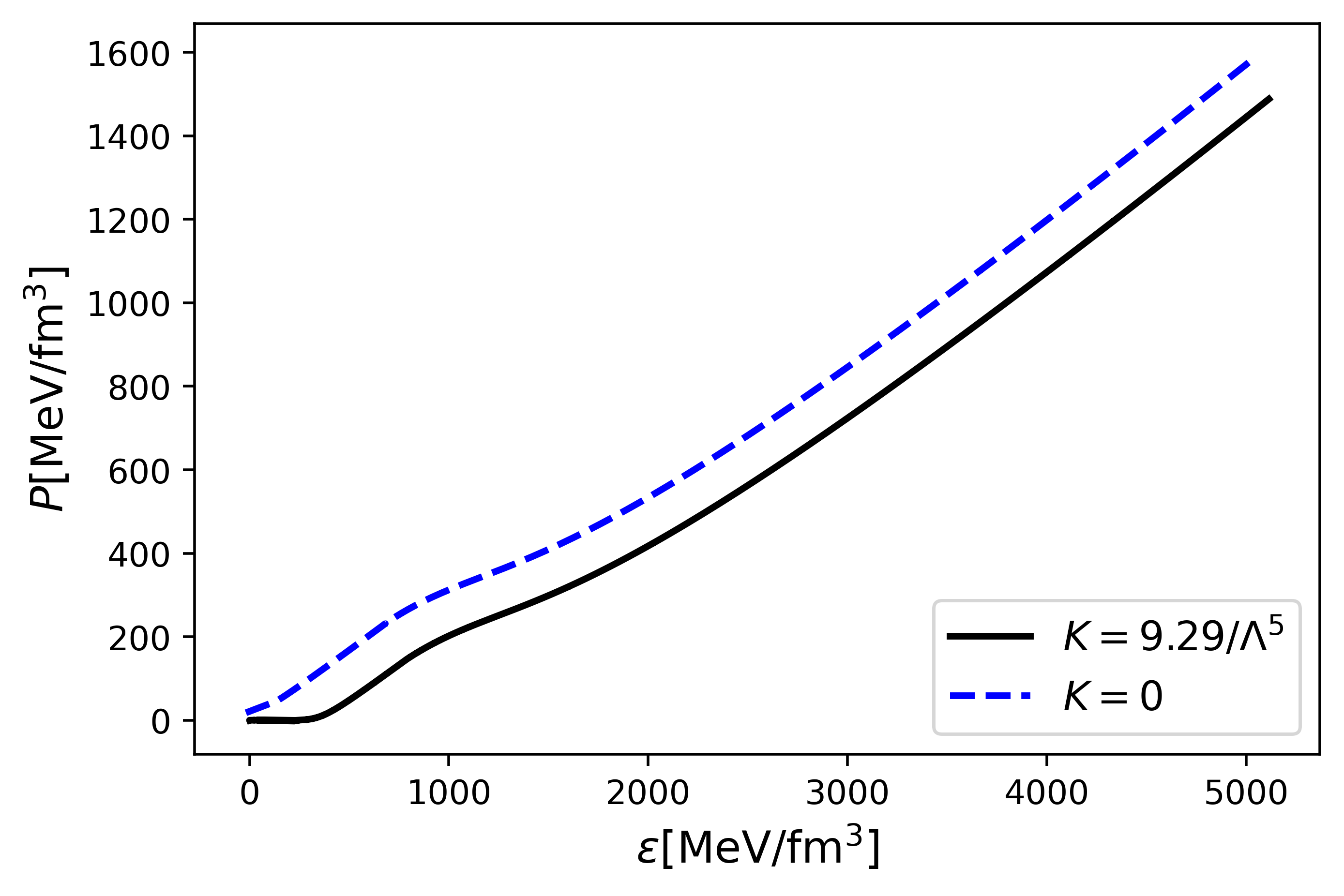}
\caption{The energy dependence of pressure for $H/G=0, g_{V}/G=0.1$.}
\label{pressure-energy}
\end{figure}

We also calculate the density dependence of the relevant pressures in Fig.\ref{pressure-density}, 
where $P_{{\rm bag}}$ is calculated from $\varepsilon_{{\rm bag}}$ using the thermodynamic relation.
\begin{align}
P_{{\rm bag}}=-\varepsilon_{{\rm bag}}+\mu_{q}n_{q}.
\end{align}
This shows that 
for same density, $P_{\text {total }} <P_{\text {total }}^{K=0}$ which is mainly caused by the difference of $P_{\rm{bag}}$,
In summary, at a given density
\begin{align}
    \varepsilon_{\text {total }} > \varepsilon_{\text {total }}^{K=0}\,,~~~~  P_{\text {total }} < P_{\text {total }}^{K=0}\,,
\end{align}
so EOS with a positive $K$ is softer, i.e., $P$ is smaller at a given $\varepsilon$, as shown in Fig.\ref{pressure-energy}.



\section{Study of properties of NS}
\label{sec:mr}

In this section,  following Ref.\cite{PhysRevC.103.045205} 
we construct a unified EOS by connecting the EOS obtained in the PDM introduced in Sec.~\ref{sec:PDM matter} 
and the EOS of NJL-type quark model given in Sec.~\ref{NJL matter}, and solve the TOV equation \cite{Tolman:1939jz,Oppenheimer:1939ne} 
to obtain the NS mass-radius  ($M$-$R$)  relation. 
As for the interplay between nuclear and quark matter EOS, see, e.g., Ref.\cite{Kojo:2020krb} for a quick review that classifies types of the interplay.


\subsection{Construction of unified EOS}

%
\begin{table}[tbh]
\begin{center}
\begin{tabular}{c|c|c|c}
\hline
\hline
$0\leq n_{B}<0.1 $ & $0.1 \leq n_{B}\leq 2n_{0}$ & $5n_{0}<n_{B}<2n_{0}$ & $n_{B}\geq 5n_{0}$\\
\hline
\rm{Crust} & \rm{PDM} & \rm{Interpolation} & \rm{NJL}\\
\hline
\hline
\end{tabular}
\end{center}
\caption{Unified EOS composed of four part}
\end{table}

In our unified equations of state,
we use the BPS (Baym-Pethick-Sutherland) EOS \cite{Baym:1971pw} as a crust EOS for $n_B \lesssim 0.5n_0$. 
From $n_B \simeq 0.5n_0$ to $2n_0$ we use our PDM model to describe a nuclear liquid.
Beyond the nuclear regime, we assume a crossover from the nuclear matter to quark matter, 
and use a smooth interpolation to construct the unified EOS. 
We expand the pressure as a fifth order polynomial of $\mu_{B}$ as
\begin{equation}
P_{\mathrm{I}}\left(\mu_{B}\right)=\sum_{i=0}^{5} C_{i} \mu_{B}^{i},
\end{equation}
where $C_{i}$  ($i=0,\cdots, 5$) are parameters  to be determined from boundary conditions  given by 
\begin{equation}
\begin{aligned}
&\left.\frac{\mathrm{d}^{n} P_{\mathrm{I}}}{\left(\mathrm{d} \mu_{B}\right)^{n}}\right|_{\mu_{B L}}=\left.\frac{\mathrm{d}^{n} P_{\mathrm{H}}}{\left(\mathrm{d} \mu_{B}\right)^{n}}\right|_{\mu_{B L}}, \\
&\left.\frac{\mathrm{d}^{n} P_{\mathrm{I}}}{\left(\mathrm{d} \mu_{B}\right)^{n}}\right|_{\mu_{B U}}=\left.\frac{\mathrm{d}^{n} P_{\mathrm{Q}}}{\left(\mathrm{d} \mu_{B}\right)^{n}}\right|_{\mu_{B U}}, \quad(n=0,1,2),
\end{aligned}
\end{equation}
with $\mu_{BL}$  being 
the chemical potential corresponding to $n_{B}=2n_{0}$ and $\mu_{BU}$ to $n_{B}=5n_{0}$. That is, we demand the matching up to the second order derivatives of pressure at each boundary.
The resultant interpolated EOS must satisfy the thermodynamic stability condition,
\beq
\chi_B = \frac{\, \partial^2 P \,}{\, (\partial \mu_B )^2 \,} \ge 0 \,,
\eeq
and the causality condition,
\begin{equation}
c_{s}^{2}=\frac{\, \mathrm{d} P \,}{\mathrm{d} \varepsilon} 
= \frac{n_{B}}{\mu_{B} \chi_{B}} \le 1 \,,
\end{equation}
which means that the sound velocity is less than the light velocity.
These conditions restrict the range of quark model parameters $(g_V, H)$ for a given nuclear EOS and a choice of $(n_L, n_U)$.

We exclude interpolated EOSs which do not satisfy the above-mentioned constraints. 
Similar surveys for the range of $(g_V, H)$ and $(n_L, n_U)$ have been carried out first
for APR EOS \cite{Akmal:1998cf} in Refs.\cite{Kojo:2014rca,Kojo:2015fua}, and more systematically
for Togashi EOS \cite{Togashi:2017mjp} in Ref.\cite{Baym:2019iky} 
and for ChEFT EOS \cite{Drischler:2020fvz} in Ref.\cite{Kojo:2021wax}.
The range explored in the present work is largely consistent with the previous works using different nuclear EOSs.
Finally we note that the estimate based on non-petrubative massive gluon exchanges
favor the estimate of $g_V \sim G$ and $H \sim 1.5G$ \cite{Song:2019qoh}.

 It is important to note that the constraints become severer for the combination of softer nucleonic EOS and stiffer quark EOS.
The rapid growth of the stiffness, together with the requirement of $c_s^2 \rightarrow 1/3$ in the high density limit,
generally leads to a peak in the sound velocity, as first found phenomenologically in Refs.\cite{Masuda:2012kf,Masuda:2012ed},
and later explained microscopically in Refs.\cite{McLerran:2018hbz,Kojo:2021ugu} with the emphasis on the quark degrees of freedom.
The growth of the stiffness in the crossover model is in general quicker than in purely hadronic models,
and such features may be studied in gravitational waves from neutron star merger events \cite{Huang:2022mqp},
or in QCD-like theories, e.g., two-color QCD, 
for which analytical \cite{Kojo:2021hqh} and lattice calculations \cite{Iida:2022hyy} suggest the rapid stiffening in the crossover domain.


%
\begin{figure*}
\centering
\subfigure[$m_{0}=400~$MeV.]{
\begin{minipage}[b]{0.48\linewidth}
\includegraphics[width=8.5cm]{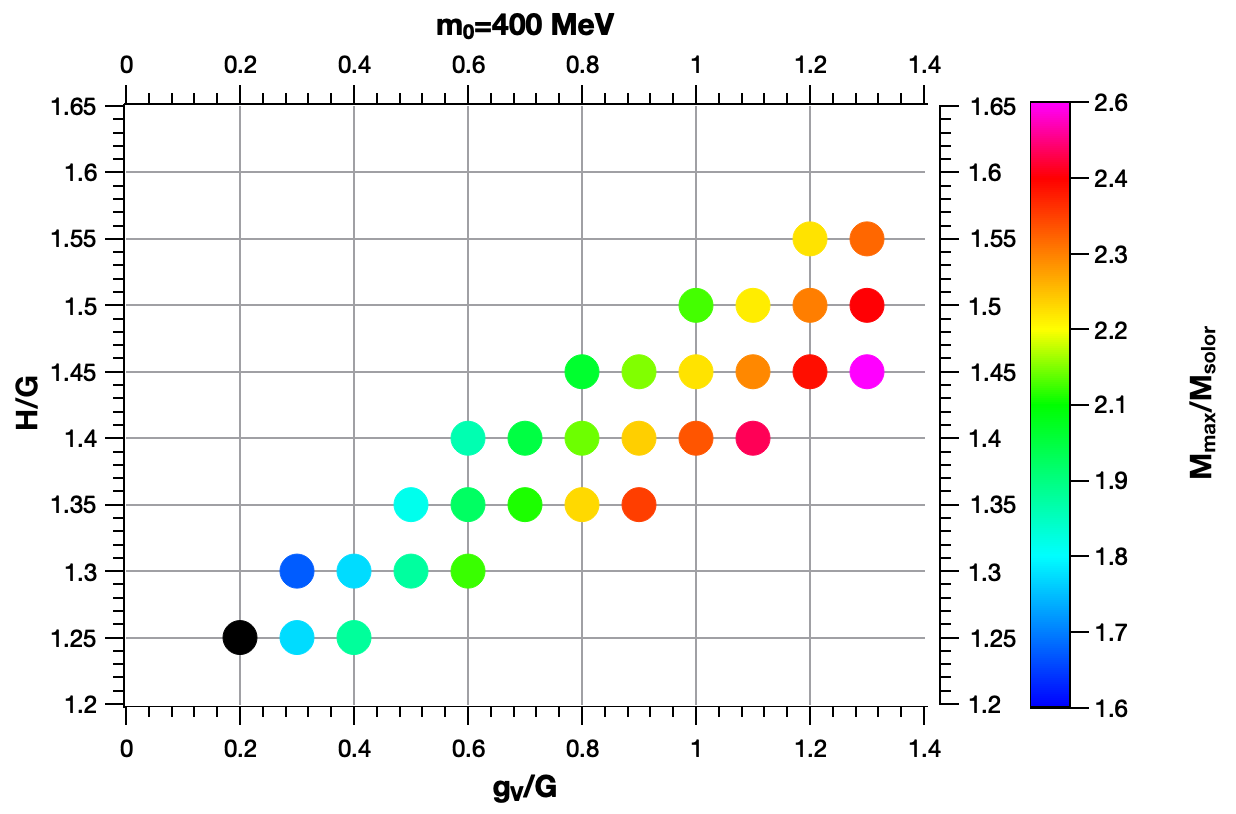}
\end{minipage}}
\subfigure[$m_{0}=500~$MeV.]{
\begin{minipage}[b]{0.48\linewidth}
\includegraphics[width=8.5cm]{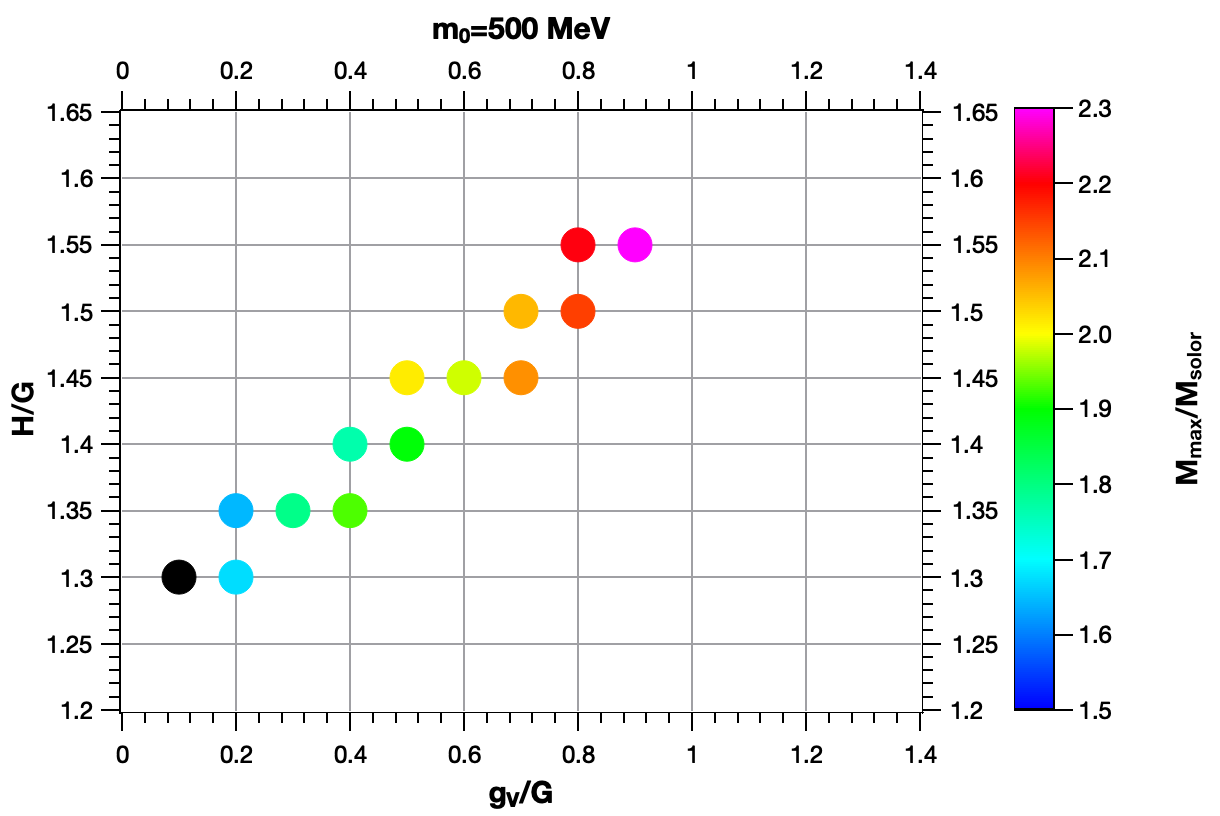}
\end{minipage}}
\subfigure[$m_{0}=600~$MeV.]{
\begin{minipage}[b]{0.48\linewidth}
\includegraphics[width=8.5cm]{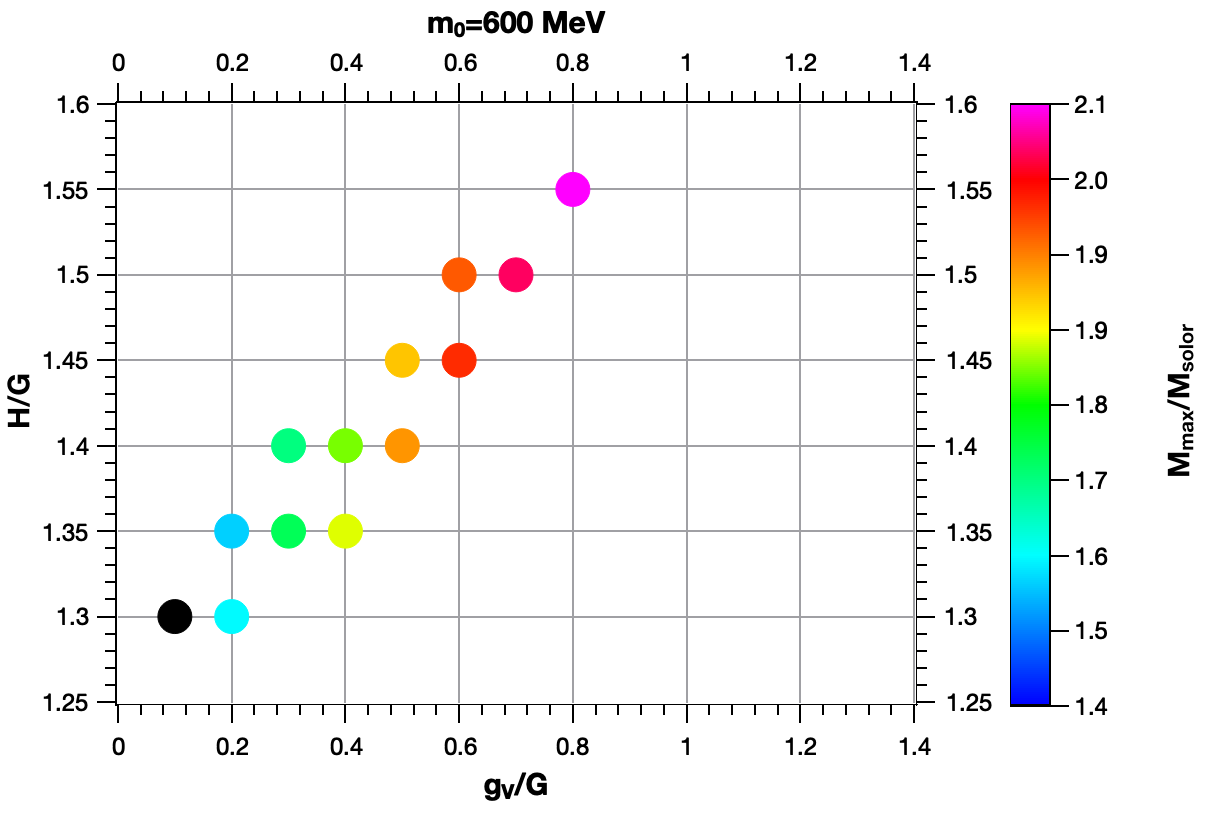}
\end{minipage}}
\subfigure[$m_{0}=700~$MeV.]{
\begin{minipage}[b]{0.48\linewidth}
\includegraphics[width=8.5cm]{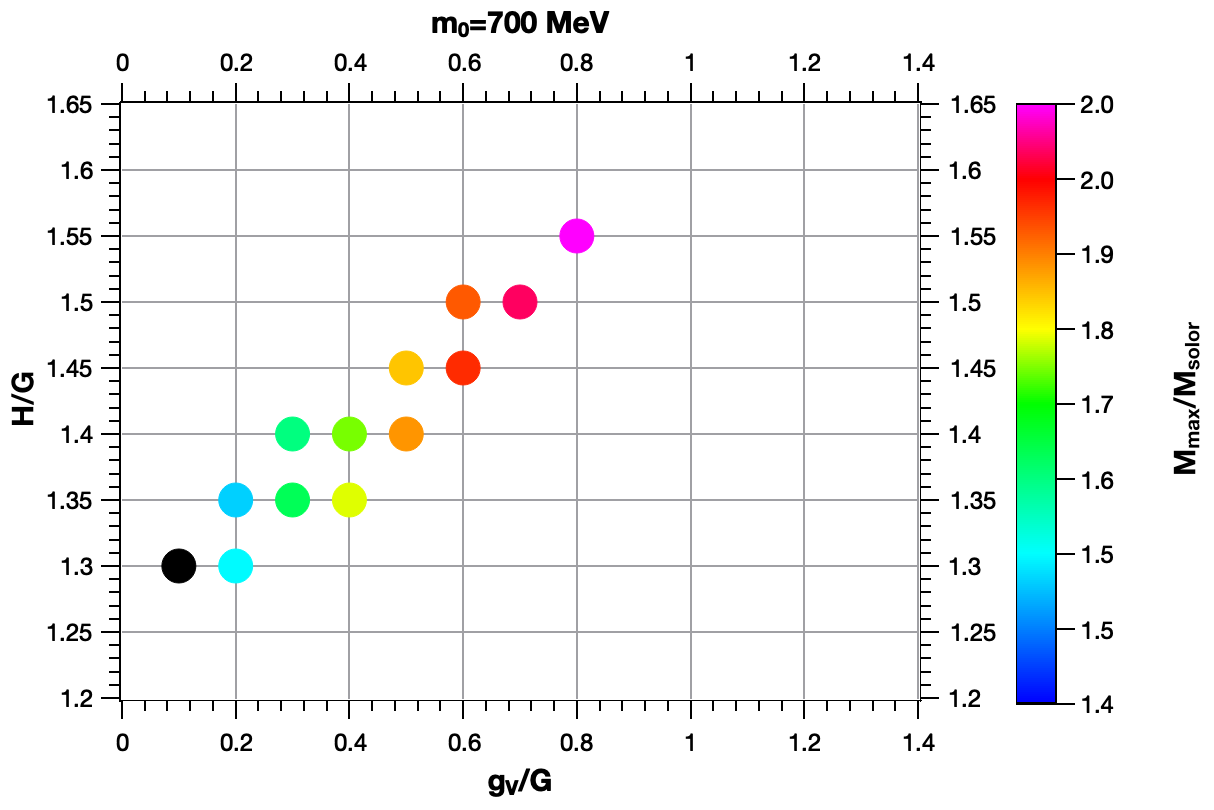}
\end{minipage}}
\subfigure[$m_{0}=800~$MeV.]{
\begin{minipage}[b]{0.48\linewidth}
\includegraphics[width=8.5cm]{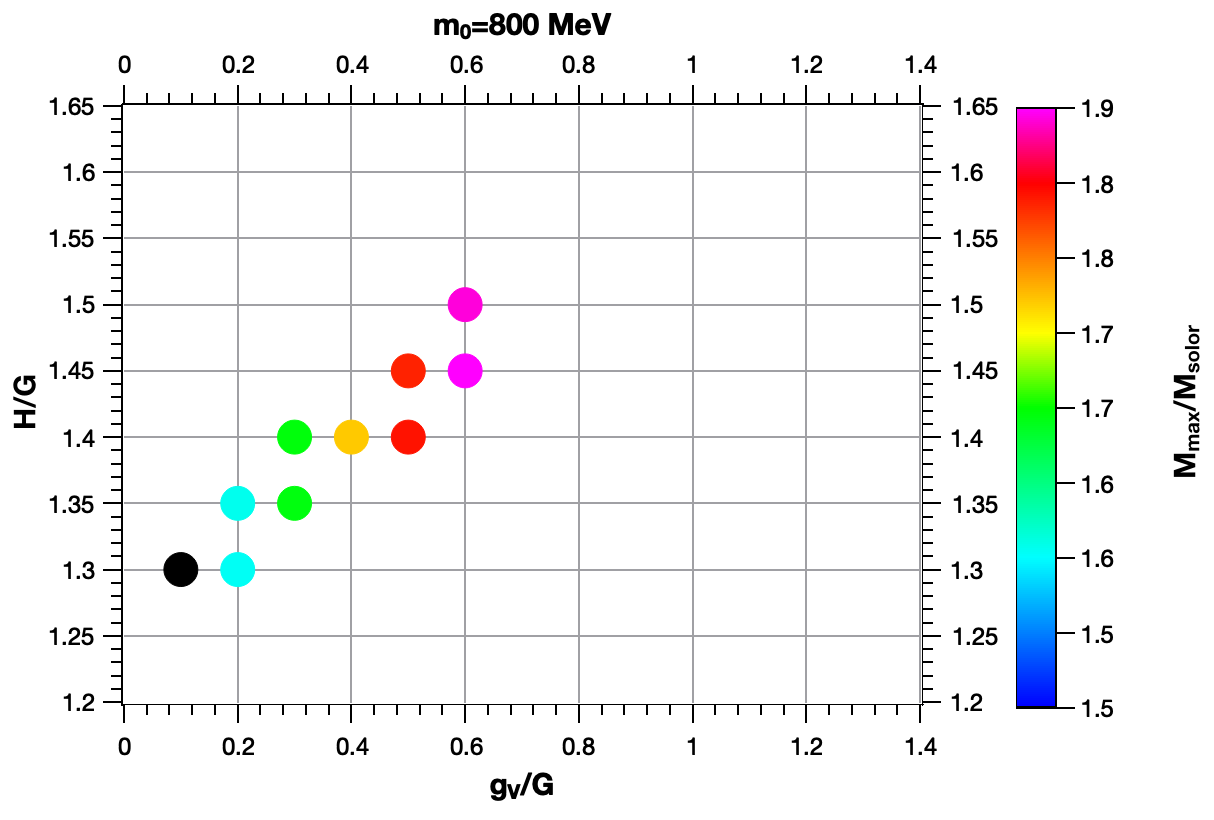}
\end{minipage}}
\caption{Allowed combinations of $(g_{V}, H)/G$ values for different $m_{0}$ choices. 
The circles indicate that the combinations are allowed and other regions are excluded by the causality condition. 
The color of the circle shows the maximum mass of NS obtained from the corresponding parameters setting.
}
\label{gvH}
\end{figure*}
%

\subsection{Mass-Radius relation}

In this section, we study  the $M$-$R$ relations of NSs 
from the unified EOS constructed above. 
In Ref.~\cite{PhysRevC.103.045205},  where the anomaly in the nuclear EOS is neglected, 
the  chiral invariant mass is constrained to be $600\, \mathrm{MeV} \lesssim m_{0} \lesssim 900\, \mathrm{MeV}$. In the present analysis, 
we improve the analyses in three aspects:
(i) we include the anomaly in the nuclear EOS; 
(ii) we newly include the $\omega^2\rho^2$-term for flexible tuning of the slope parameter $L$ in the symmetry energy (here we adopt the value $L=57.7$\,MeV as a baseline suggested by Ref.~\cite{universe7060182});
and (iii) we include a new constraint from the NICER on the radius of $2.1M_\odot$ neutron stars.

We first examine the effects of NJL parameters ($g_{V}, H$). 
For simplicity, we fix parameters in the PDM to $B=600~$MeV, $\lambda^{'}_{8}=0, \lambda^{'}_{10}=0.44$,  
and $\lambda_{\omega \rho}$ tuned to reproduce $L=57.7~$MeV.
We then vary the value of $m_0$ and examine the range of $(g_V,H)$ which is allowed by the causality and thermodynamic stability conditions.
The band shown in the Fig.\ref{gvH} specifies such domains, while the blank part is not allowed.
A larger $g_V$ requires a larger $H$.
For $m_{0}=800~$MeV, the maximum masses for all the combinations are below 2$M_{\odot}$,
leading to the conclusion that $m_{0}=800~$MeV should be excluded within the current setup of the PDM parameters.

\begin{figure}
\centering
\includegraphics[width=8cm]{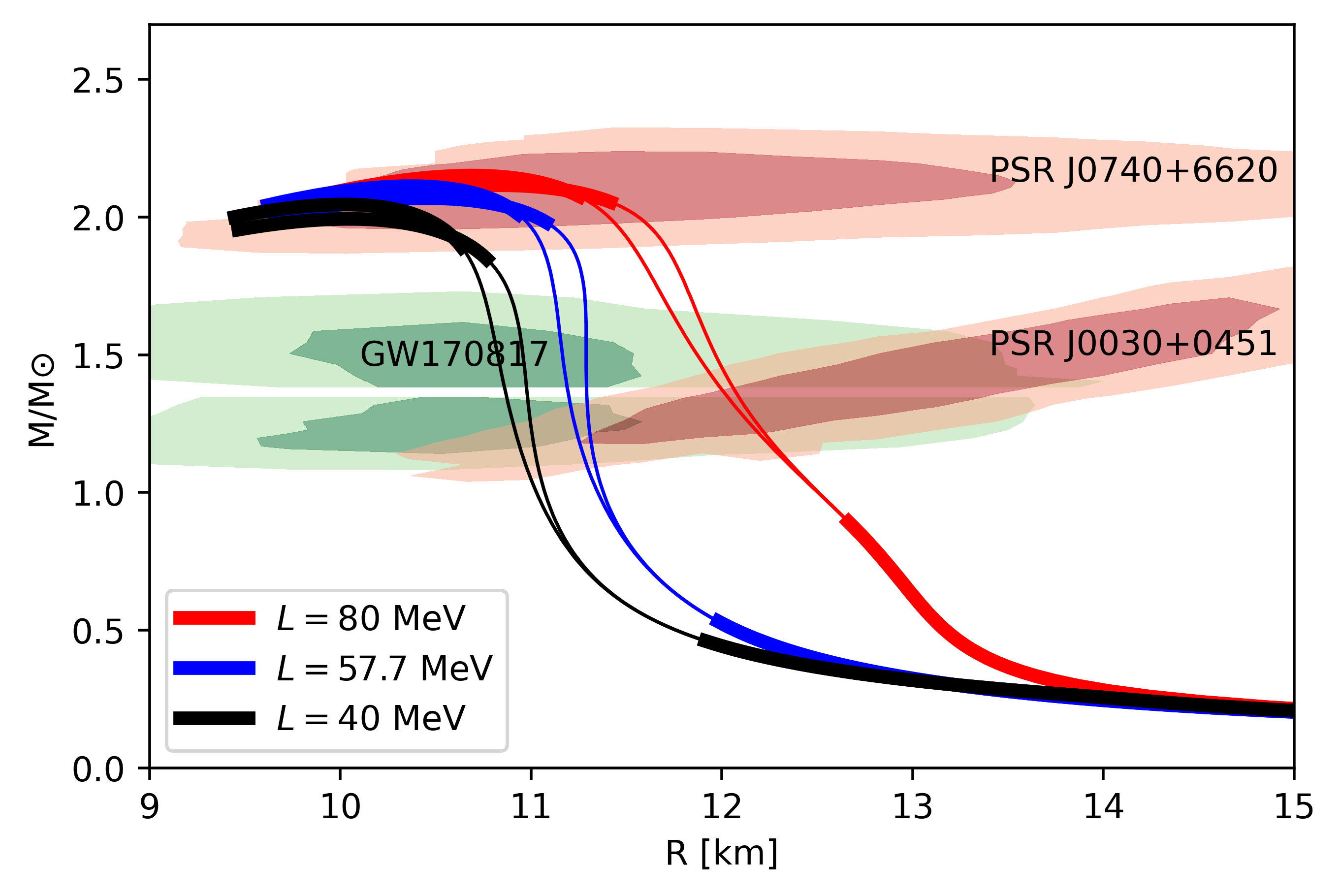}
\caption{Mass-radius relations for  $m_{0}=500~$MeV with different slope parameter. 
Red curves are connected to the NJL parameters $(H, g_V)/G=$ (1.55, 1.0), (1.5, 0.9);
blue curves to (1.55, 0.9), (1.5, 0.8); black curves to (1.55, 0.8), (1.5, 0.7).
}
\label{gv_H}
\end{figure}

Next we fix $m_0=500$ MeV and vary the value of $\lambda_{\omega \rho}$ or $L$ while the rest of hadronic parameters kept unchanged. 
The resultant $M$-$R$ relation is shown in Fig.~\ref{gv_H}, 
thick curves  in the low (high)-mass region indicate the central density of the NS is smaller than 2$n_0$ (larger than 5$n_0$), and the NS is made from hadronic matter (quark matter).  
The thin curves on the other hand show that the core is in crossover region. 
From the figure one sees that the EOSs are softened by the effect of the $\omega^2\rho^2$-term 
and the radius for $L=57.7~{\rm MeV}, M \simeq 1.4 M_\odot$ is about $11.2~$km in comparison with the result of $L=80~$MeV about $12.1~$km.  
There is still a large ambiguity about the values of slope parameter and small slope parameters usually soften the NS EOS, shifting the radius towards smaller values. 
Precise determination of slope parameter in the future will help us further constrain the NS properties.

\begin{figure*}
\centering
\subfigure[$B=0, 600~$MeV for $m_0=500, 800~$MeV.]{
\begin{minipage}[b]{0.48\linewidth}
\includegraphics[width=8cm]{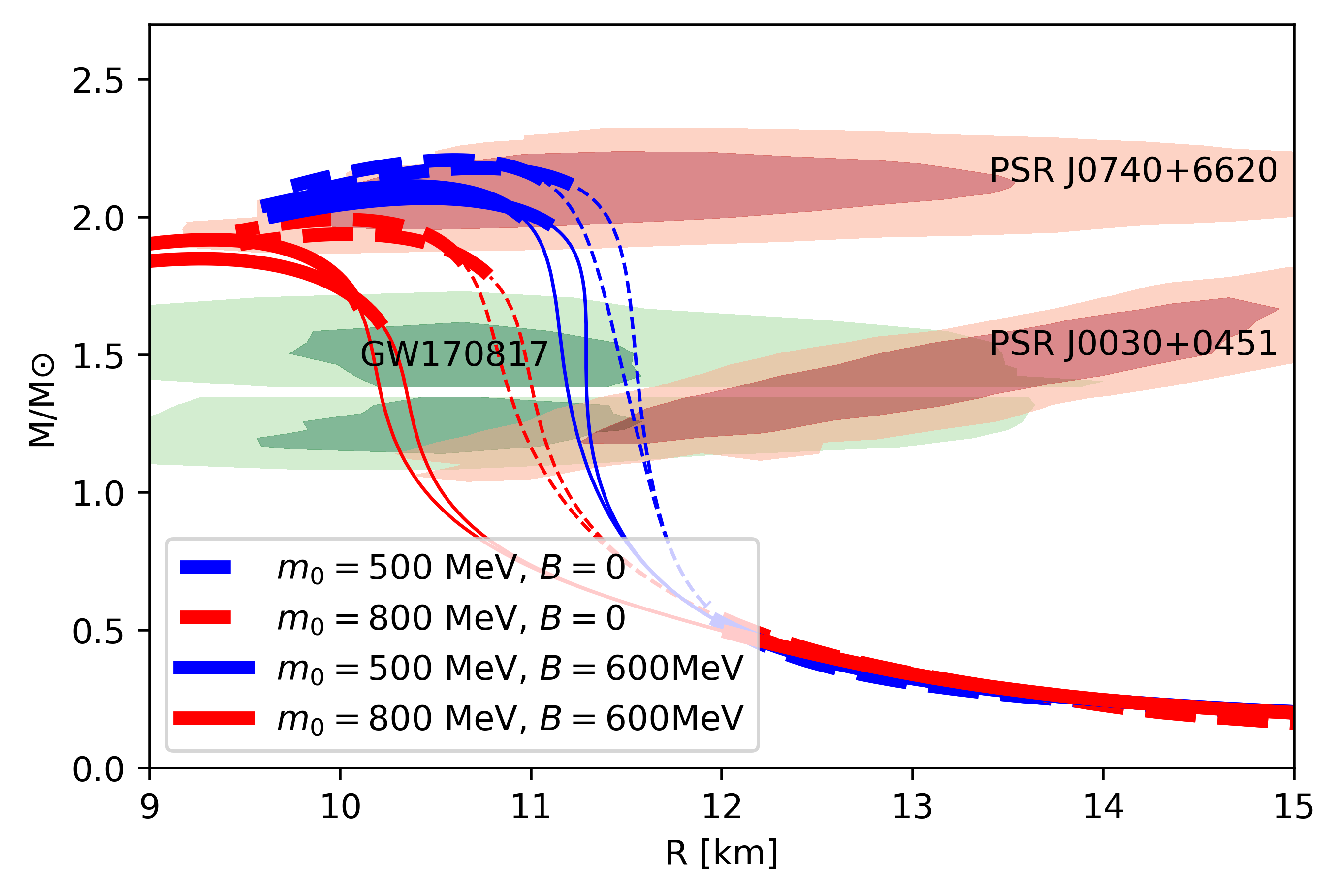}
\end{minipage}}
\subfigure[$B=600~$MeV for different $m_{0}$. NJL parameters $(H, g_V)/G$ are chosen to be (1.45,1.3)$_{m_{0}=400{\rm MeV}}$, (1.6,1.3)$_{m_{0}=500{\rm MeV}}$, (1.6,1.3)$_{m_{0}=600{\rm MeV}}$, (1.6,1.2)$_{m_{0}=700{\rm MeV}}$.]{
\begin{minipage}[b]{0.48\linewidth}
\includegraphics[width=8cm]{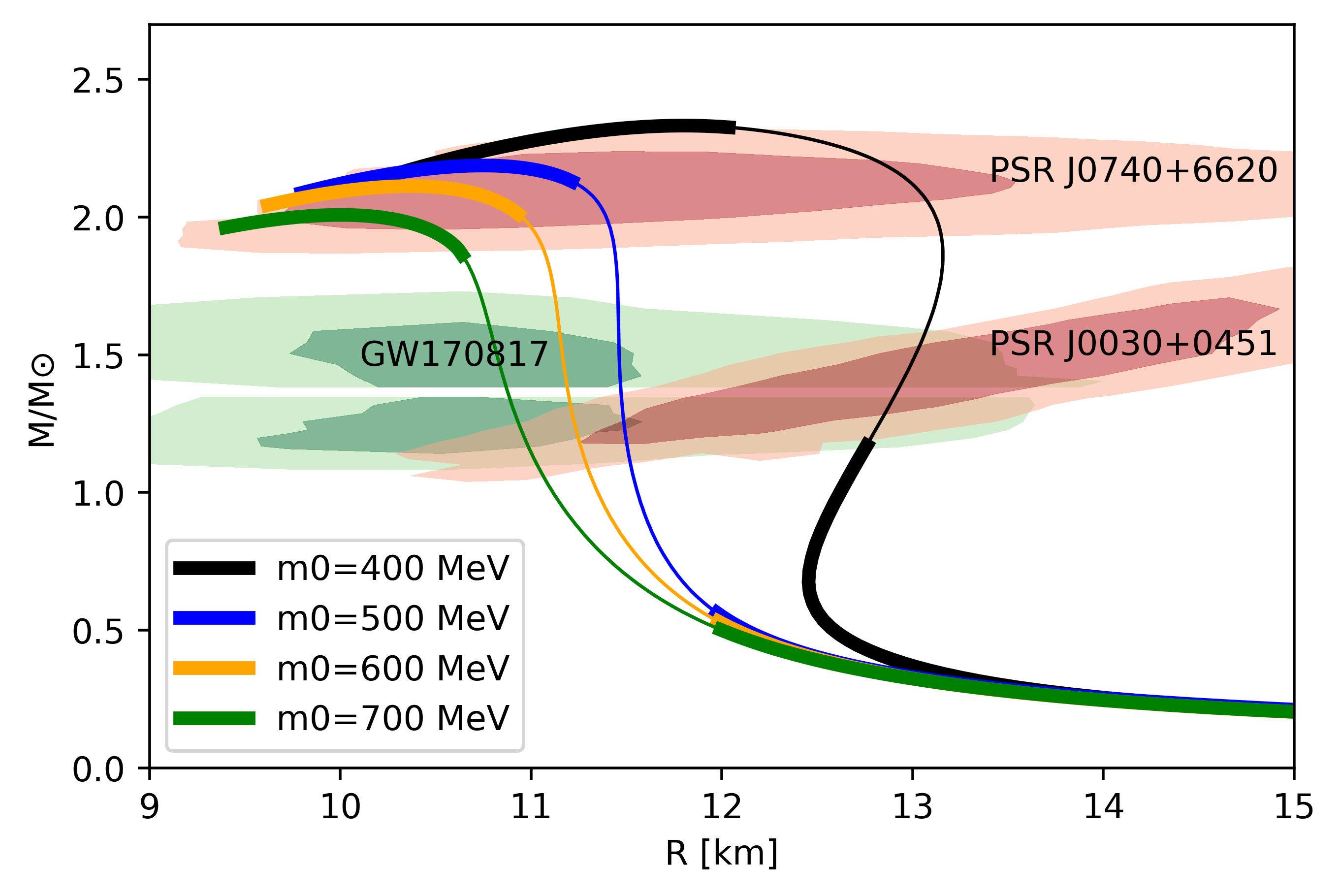}
\end{minipage}}
\caption{Mass-radius relations for different $m_{0}$ in different parameter setting. 
}
\label{mass-radius}
\end{figure*}

In following analysis, we fix the value $L=57.7~$MeV and examine the effects of anomaly on the $M$-$R$ relation. 
In Fig. \ref{mass-radius}(a), we  show how the $M$-$R$ curves change under the $B$ effect.
The ($\lambda^{'}_{8}, \lambda^{'}_{10}$) parameters from  $m_{0}=400$ to $800~$MeV are fixed to the boundary values in the following analysis, $\lambda_{8}^{'}=0, \lambda_{10}^{'}=0.44$.
The NJL parameter $(H, g_{V})$  are chosen to have the stiffest two $M$-$R$ curves. In the Fig. \ref{mass-radius}(a), because of the softening effect of anomaly, after we set $B=600~$MeV, the stiffest connection for $m_{0}=800~$MeV is unable to satisfy the maximum constraints. 
In Fig. \ref{mass-radius}(b), we show the final results in this work after setting $B=600~$MeV for different $m_0$ values. 
We find the final constraints to the chiral invariant mass is changed to be smaller by $\sim 100$ MeV 
in comparison with the previous constraints in Ref.~\cite{PhysRevC.103.045205}.

%


\section{A Summary and discussion}
\label{sec:summary}

In this work, we  construct an  effective hadronic model
in which the effect of strange quark condensate is included in the mesonic sector through 
the  Kobayashi-Maskawa-'t\,Hooft (KMT)-type interaction  reflecting  the $U(1)$ axial anomaly.
We then study the impact of $U(1)_{A}$ anomaly on the chiral symmetry breaking in both hadronic and an NJL-type quark modes. 
 In both models the $U(1)_{A}$ anomaly enhances the chiral symmetry breaking. 
In the PDM, the anomaly effects increases the effective mass of $\sigma$,
and the heavier $\sigma$ meson mass reduces the range of attractive force, weakening the overall strength;
this in turn requires weaker repulsive $\omega$ interactions to balance with the $\sigma$ attraction to satisfy the saturation properties.
The resultant reduced repulsion leads to a softer nuclear EOS at supra-saturation densities.
In the  NJL-type model, the anomaly effects lead to large bag constant. 
Since a larger bag constant adds the energy density but reduces the pressure, the corresponding EOS is softened. 
We expect that it is a general feature that $U(1)_{A}$ anomaly softens the NS EOS.

The EOS plays an essential role when determining the NS properties.
The NICER analyses of the most massive NS known, PSR J0740+6620, with $M/M_{\odot}= 2.08\pm0.07$
and the radii $R_{2.08} =12.35 \pm 0.75$ km \cite{Miller:2021qha},
together with the updated estimate for  $R_{1.4} =12.35 \pm 0.75$ km \cite{Miller:2021qha},
disfavors strong first order phase transitions in the region between $1.4M_\odot$ and $2.1M_\odot$. 

In this case, we assume the hadronic and quark matter are not distinctly different and  construct unified EOS for neutron star matter.  
At present work, we interpolate the EOS
obtained in the hadronic model based on the parity doublet structure ($n_{B} \leq 2 n_{0}$) 
and the one in the NJL-type quark model ($n_B \ge 5n_{0} $) with crossover in the intermediate region. 
We found that the unified  EOS is also softened by the effect of anomaly 
 due to the softening of the EOS in both hadronic and quark matters.  
The resultant $M$-$R$ curves are compared with the constraints  
from GW170817 (LIGO $\&$ VIRGO) and PRS J0030+0451 (NICER) as well as the constraint from PRS J0740+6620.
 From the constraints  we restrict the chiral invariant mass as 
\begin{align}
    400\, {\rm MeV} \lesssim m_{0} \lesssim 700\, {\rm MeV} \,.
\end{align}
Compared with results without anomaly, $500\, {\rm MeV} \lesssim m_{0} \lesssim 800\, {\rm MeV}$, 
we find that the anomaly softens the EOS, shifting the range of chiral invariant mass towards lower values by $100~$MeV. 
The effects of the $\omega^{2}\rho^{2}$ term or $L$ are also very important when we constrain the chiral invariant mass as shown in Fig.~\ref{gv_H}.
Small values of $L$ not only decrease the total radius but also lead to smaller maximum mass. 
The typical estimates value is $L=30$-80 MeV \cite{Tews:2016jhi,Drischler:2020hwi},
but recent PREXII for the neutron skin thickness suggested $L=109.56 \pm 36.41~$MeV \cite{Reed:2021nqk}.
Future determination of $L$ from the experiments will help us to better constrain the values of chiral invariant mass.

In this paper, we included the anomaly $B$ term only in the mesonic sector. 
We may include some Yukawa interactions which also break the $U(1)_A$ symmetry. 
Furthermore,  we can add hyperons to a hadronic model with parity doublet structure based on 
the $SU(3)_L \otimes  SU(3)_R$ chiral symmetry combined with the $U(1)_A$ anomaly.
We leave these analyses as future works.
 
 \begin{acknowledgments}

The work of B.G., T.M., and M.H. was supported in part by JSPS KAKENHI Grant No. 20K03927. T.M. was also supported by JST SPRING, Grant No. JPMJSP2125.; T.K. by NSFC grant No. 11875144 and by the Graduate Program on Physics for the Universe at Tohoku university.
\end{acknowledgments}

\appendix

\section{CHIRAL CONDENSATES}
We calculate the chiral condensates in the PDM by differentiating the thermodynamic potential with respect to the current quark masses\cite{Minamikawa:2021fln}:
\begin{align}
\langle(\bar{u} u+\bar{d} d)\rangle &\equiv \frac{\partial \Omega_{\mathrm{PDM}}}{\partial m_{q}}.  
\end{align}
Then, using the Gell-Mann-Oakes-Renner relation, we obtain
\begin{align}
\frac{\langle(\bar{u} u+\bar{d} d)\rangle}{\langle(\bar{u} u+\bar{d} d)\rangle_{0}}&=\frac{\sigma}{f_{\pi}}.
\end{align}
Similarly, we obtain
\begin{align}
\langle\bar{s}s\rangle \equiv \frac{\partial \Omega_{{\rm PDM}}}{\partial m_{s}},
\end{align}
and 
\begin{align}
\frac{\langle\bar{s}s\rangle}{\langle\bar{s}s\rangle_{0}}=\frac{\sigma_{s}}{\sigma_{s0}}.
\end{align}
where $\sigma_{s0}$ is the mean field $\sigma_{s}$ at the vacuum.

\begin{figure*}
\centering
\subfigure[]
{
\begin{minipage}[b]{0.48\linewidth}
\includegraphics[width=8cm]{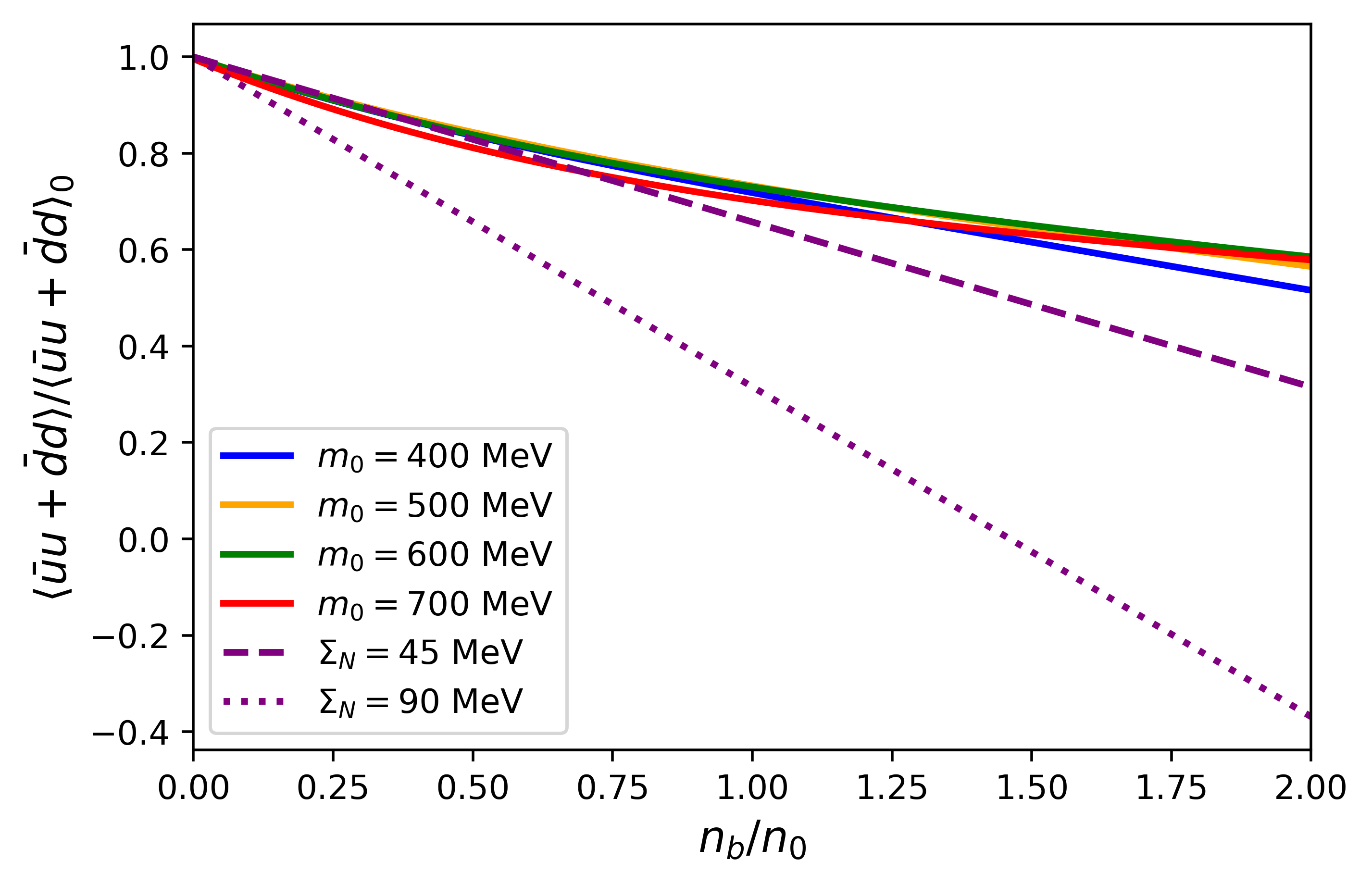}
\end{minipage}}
\subfigure[]{
\begin{minipage}[b]{0.48\linewidth}
\includegraphics[width=8cm]{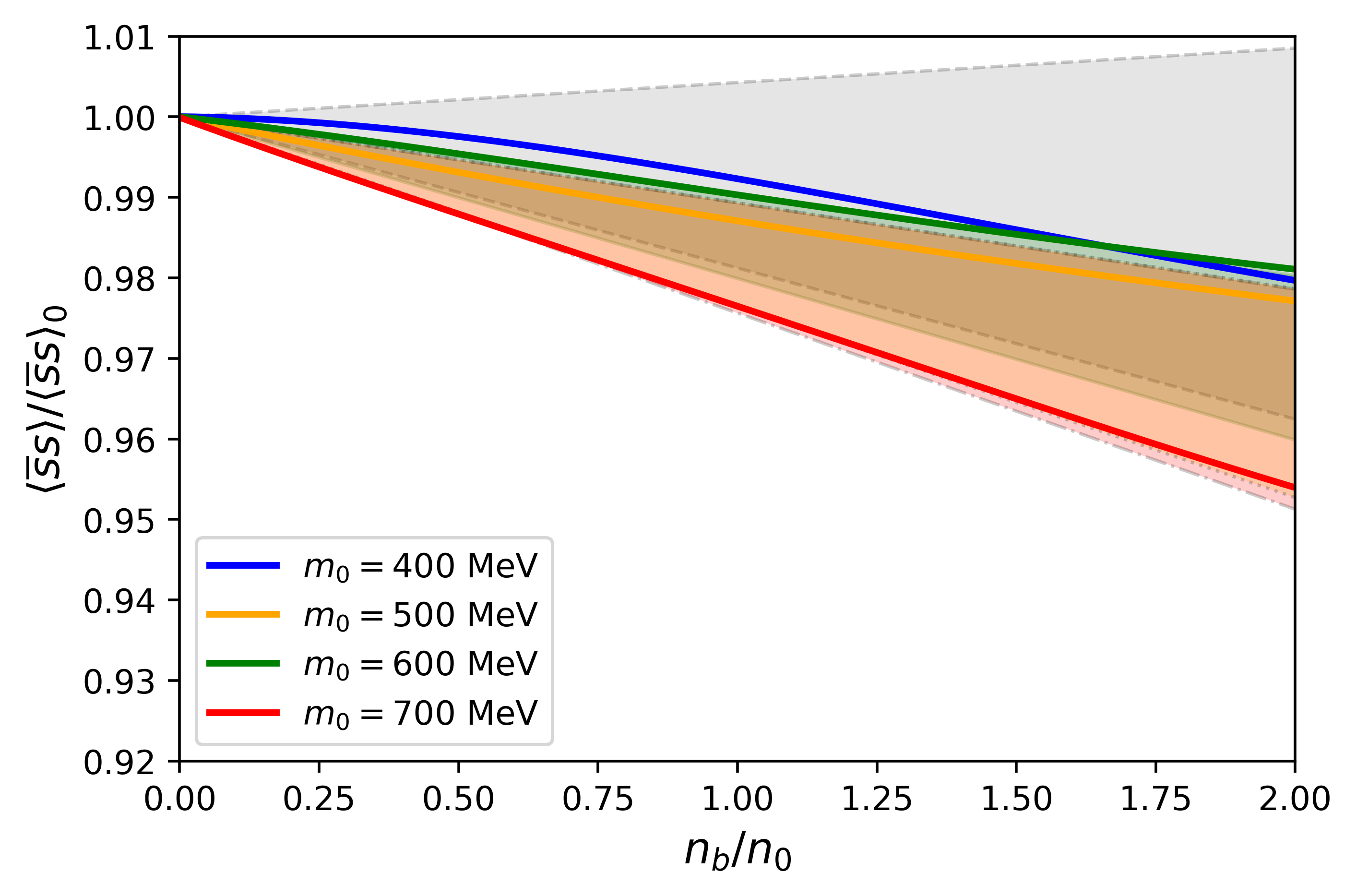}
\end{minipage}}
\caption{Density dependence of the chiral condensate $\langle \bar{u}u + \bar{d}d \rangle/ \langle \bar{u}u + \bar{d}d\rangle_{0}$ (left)  and $\langle \bar{s}s \rangle/\langle \bar{s}s \rangle_{0}$ (right) in the PDM.  We use the same parameter choices in Fig. \ref{mass-radius}(b). In (a), two dotted lines show  the density dependence in the linear density approximation with $\Sigma_{N}=45,90~$MeV as typical examples. In (b), the colored bands are drawn in the linear density approximation where  the value of  $\Sigma_{sN}$ is taken from the lattice QCD results shown in Table.\ref{lattice} with error bars included: JLQCD(grey band), RQCD(green band), ETM(red band) and  $\chi$QCD(orange band).
}
\label{fig:condensate}
\end{figure*}

 In the linear density approximation, the density dependence of the condensates are given by 
\begin{align}
&\langle\bar{q} q\rangle \simeq\langle 0|\bar{q} q| 0\rangle+\rho\langle N|\bar{q} q| N\rangle=\langle 0|\bar{q} q| 0\rangle+\rho \frac{\Sigma_{ N}}{2 m_{q}}, \\
&\langle\bar{s} s\rangle \simeq\langle 0|\bar{s} s| 0\rangle+\rho\langle N|\bar{s} s| N\rangle=\langle 0|\bar{s} s| 0\rangle+\rho \frac{\Sigma_{s N}}{m_{s}},
\end{align}
where $\Sigma_{N}$ is the $\pi N$ sigma term and $\Sigma_{sN}$ is the strange quark sigma term. 

In Fig.\ref{fig:condensate}(a), we show the density dependence of $\langle(\bar{u} u+\bar{d} d)\rangle/\langle(\bar{u} u+\bar{d} d)\rangle_{0}$ determined from the PDM in the neutron star matter. We also plot typical examples of the density dependence of the condensate determined in the linear density approximation where the $\pi N$ sigma term is taken as $\Sigma_{N}=45, 90~$MeV Ref.\cite{Minamikawa:2021fln}.  This shows that in the low density region, the density dependence of  chiral condensate obtained in our model is consistent with the linear density approximation, while there is some deviation in density region $n_{b}/n_{0}\gtrsim 0.5$ due to the higher order correction.

In Fig.\ref{fig:condensate}(b), we show the density dependence of strange quark chiral condensate compared with the linear density approximation shown by colored bands. 
In the linear density approximation, 
we use the value of $\Sigma_{sN}$ determined by the  lattice QCD simulations shown in Table.\ref{lattice} as  typical examples.
The colored bands  in Fig.15(b) are written by taking account of all the errors, e.g. $\Sigma_{sN}=40.2 \pm 15.2~$MeV for $\chi$QCD\cite{Yang:2015uis}.
This Fig.\ref{fig:condensate}(b) shows that the ambiguity of  $\Sigma_{sN}$ is too large to give a constraint to our model. However,  we expect that the precise determination of $\Sigma_{sN}$ in future will   constrain the chiral invariant mass.

\begin{table}[tbh]
\begin{center}
\begin{tabular}{c|c}
\hline Collaboration & $\Sigma_{s N}[\mathrm{MeV}]$  \\
\hline  
 $\chi \mathrm{QCD}$ & $40.2(11.7)(3.5)$\cite{Yang:2015uis} \\
 ETM & $41.1(8.2)(7.8)$\cite{Abdel-Rehim:2016won}  \\
 RQCD & $35(12)$\cite{Bali:2016lvx} \\
 JLQCD & $17(18)(9)$\cite{Yamanaka:2018uud} \\
\hline
\end{tabular}
\end{center}
\caption{Values of  $\Sigma_{sN}$ obtained by recent from Lattice QCD simulations.}
\label{lattice}
\end{table}


\newpage

\bibliographystyle{unsrt}
\bibliography{ref.bib}

\end{document}